\documentclass[12pt,preprint]{aastex}
\usepackage{epsfig}
\shorttitle{IR Dust Bubbles}
\shortauthors{Watson et al.}
\begin{document}
\title{IR Dust Bubbles: Probing the Detailed Structure and Young Massive
  Stellar Populations of Galactic HII Regions} \author{C.
  Watson\altaffilmark{1}, M.S. Povich\altaffilmark{2}, E.B.
  Churchwell\altaffilmark{2}, B.L. Babler\altaffilmark{2}, G.
  Chunev\altaffilmark{1}, M.  Hoare\altaffilmark{5}, R.
  Indebetouw\altaffilmark{3}, M.R.  Meade\altaffilmark{2}, T.P.
  Robitaille\altaffilmark{6}, B.A.  Whitney\altaffilmark{4}}
\altaffiltext{1}{Manchester College, Dept. of Physics, 604 E. College Ave.,
  North Manchester, IN 46962} \altaffiltext{2}{Univ. of Wisconsin - Madison,
  Dept. of Astronomy, 475 N. Charter St., Madison, WI 53716}
\altaffiltext{3}{Department of Astronomy, University of Virginia, P.O. Box
  3818, Charlottesville, VA 22903-0818} \altaffiltext{4}{University of
  Colorado, Space Science Institute, 1540 30th St., Suite 23, Boulder, CO
  80303-1012} \altaffiltext{5}{School of Physics and Astronomy, University of
  Leeds, Leeds, West Yorkshire, LS2, 9JT} \altaffiltext{6}{5SUPA, School of
  Physics and Astronomy, University of St Andrews, North Haugh, KY16 9SS, St
  Andrews, United Kingdom}
\begin{abstract}
  We present an analysis of wind-blown, parsec-sized, mid-infrared bubbles and
  associated star-formation using GLIMPSE/IRAC, MIPSGAL/MIPS and MAGPIS/VLA
  surveys.  Three bubbles from the Churchwell et al. (2006) catalog were
  selected. The relative distribution of the ionized gas (based on 20 cm
  emission), PAH emission (based on 8 $\mu$m, 5.8 $\mu$m and lack of 4.5
  $\mu$m emission) and hot dust (24 $\mu$m emission) are compared. At the
  center of each bubble there is a region containing ionized gas and hot dust,
  surrounded by PAHs. We identify the likely source(s) of the stellar wind and
  ionizing flux producing each bubble based upon SED fitting to numerical hot
  stellar photosphere models.  Candidate YSOs are also identified using SED
  fitting, including several sites of possible triggered star formation.
\end{abstract}
\section{Introduction}
Mid-infrared (MIR) imaging surveys of the Galactic plane such as the
Mid-Course Space Experiment (MSX; Price et al.  2001) and the Galactic Legacy
Infrared Midplane Survey Extraordinaire (GLIMPSE; Churchwell et al. 2001;
Benjamin et al. 2003) have revealed a large number of bubbles in the disk of
the Galaxy.  The bubbles have bright 8 $\mu$m shells that enclose bright 24
$\mu$m interiors.  ISO (Infrared Space Observatory; Kessler et al.  1996) and
MSX were the first to reveal the existence of these objects (including the
association with triggered star-formation, Deharveng et al.  2005), but the
GLIMPSE survey (with ten times better spatial resolution and a hundred times
the sensitivity than MSX) has enabled the detection of many more bubbles and
has enabled the refinement of their morphological structure.  In fact, the
GLIMPSE survey has shown that bubbles are a hallmark of the diffuse emission
in the Galactic plane.  Small and faint bubbles are apparent only in the
GLIMPSE survey due to the lower resolution and sensitivity of MSX.  Churchwell
et al.  (2006; 2007) have cataloged almost 600 bubbles. An average of
$\gtrsim$5 bubbles per square degree were found within 20$^\circ$ of the
Galactic center.  These bubbles are identified as complete or partial rings in
the GLIMPSE images, which Churchwell et al.  (2006) argue are two dimensional
representations of three dimensional bubbles.  Churchwell et al. (2006) showed
that the bubbles are distributed in longitude and latitude like O and B stars,
$>$25\% of which are coincident with known radio HII regions, and $\sim$13\%
enclose known stellar clusters.  About 1/3 of the bubbles are produced by O
stars.  Approximately two thirds of the sample have small angular diameters,
which Churchwell et al.  (2006) propose to be mostly physically small bubbles
produced by late B type stars whose UV photon fluxes are adequate to excite a
shell of PAH emission but not adequate to produce a detectable radio HII
region.  Churchwell et al.  (2006; 2007) tabulated bubble morphological
properties such as angular diameter, shell thickness, and eccentricity.


Bubbles/HII regions identified by MIR emission present a new and powerful tool
to study the interaction of young hot stars with their environment.  Young
stars impact the ambient interstellar medium (ISM) by heating the dust,
ionizing the gas (in the case of O and early B stars), and heating an
expanding bubble.  Bubbles can be produced by stellar winds and over-pressure
by ionization and heating by stellar UV radiation.  Bubble sizes and
morphologies are determined by the density structure of the ambient ISM, the
luminosity of the central star(s) responsible for producing the bubble, the
stellar wind luminosity and the motion of the central star(s) relative to the
ambient ISM.  The morphologies of MIR dust bubbles/HII regions reveal
important information about the strength and directionalities of stellar winds
and the structure and physical properties of the ambient ISM into which the
bubbles are expanding.  Pillars, scalloping, and sharpness of the inner
boundaries of bubbles defined by 8$\mu$m emission provide unique insights into
the hydrodynamics, photo-ionization, and evaporation of gas and sublimation of
dust in expanding bubbles, and stellar mass loss rates during their evolution.
For example, based on GLIMPSE-MIR morphology of ultra-compact (UC) HII
regions, Hoare et al.  (2007) and references therein have found an unusually
high fraction of cometary shapes. They argue that the morphology of the
PAH-traced photo-dissociated region (PDR), combined with the radio free-free
emission, is suggestive of a champagne flow.

MIR bubbles that surround HII regions ionized by O stars and B supergiants
(i.e. those that have strong stellar winds) have been modeled by several
groups.  Analytic evolution models have been developed by Castro, McCray, and
Weaver (1975), Weaver et al. (1977), Whitworth et al. (1994), and Capriotti
and Kozminsky (2001). These models predict that wind-blown bubbles around O
stars and B supergiants should have the following distinct regions: a
hypersonic stellar-wind-evacuated central cavity surrounding the central star
where densities are very low; a hot, low-density, shocked stellar wind region
surrounding the central evacuated cavity that occupies most of the volume of
the HII region; a thin conduction zone where temperature sharply decreases and
density rapidly increases (with radius); a thin, dense shell of shocked H$^+$
gas surrounded by a very thin shell of non-shocked H$^+$ gas at about 10$^4$
K. The outer thin shell of 10$^4$ K H$^+$ gas is the region classically
thought to represent the photo-ionized HII region. This basic picture is
supported by detailed numerical evolution models calculated for a 35 M$_\odot$
(Freyer, Hensler, \& Yorke, 2006; hereafter FHY06), a 60 M$_\odot$ (Freyer,
Henseler, and Yorke, 2003), and an 85 M$_\odot$ star (Kr\"oger, Hensler, \&
Freyer, 2006, and Kr\"oger, Freyer, Hensler, \& Yorke, 2007).  The numerical
models, however, show that the sharp boundaries predicted by the analytic
models probably are not sharp due to turbulent mixing.  The consequence of
this is that clumps of cool 10$^4$ K gas are predicted to be found mixed with
the hot 10$^7$ K gas.  The thickness of the 10$^4$ K shell is predicted to be
a function both of age and luminosity of the central star; the lower the wind
luminosity the thicker the cool ionized shell is.  Also the outer boundaries
of the HII regions are predicted to be quite jagged due to instabilities
produced when a dense medium expands into a much lower density interstellar
medium (ISM).  We will compare some of the model predictions with our
observations in \S 5. All theoretical evolution models up to now have omitted
the possible role of dust in HII regions, which new MIR observations clearly
demonstrate is present.

Recent observations are providing a deeper understanding of the relative
distributions of thermal dust, PAHs, ionized gas, and PDRs.  Peeters et al.
(2005) review the ISO spectroscopic observations of compact and evolved HII
regions and the PDRs surrounding them. They find gas temperatures in the PDRs
between 100 K (NGC 2024) and 200 K - 500 K (S106IR) and densities between
10$^4$ cm$^{-3}$ (W75N) and 10$^6$ cm$^{-3}$ (NGC 2024).  They find broad
agreement between these results and models of the illumination of PDRs by the
appropriate nearby hot star(s). Kassis et al. (2006), on the other hand,
report ground-based observations of the Orion bar that indicate PAHs are
present just outside the HII region. They show that [Ne II] 12.81 $\mu$m
emission traces the geometry of the PDR in the bar. They also conclude that
there is not a sharp transition between neutral and ionized PAH emission
within the PDR.

Each of the four Spitzer/IRAC (Infrared Array Camera) bandpasses are dominated
by different emission processes in the neighborhood of a hot star (see Draine
2003 and Peeters et al. 2003).  The brightest objects in the 3.6$\mu$m band
are stars, but this band also has contributions from a weak, diffuse PAH
feature at 3.3$\mu$m, and possibly from scattered star light.  The 4.5$\mu$m
band samples diffuse emission in the Br$\alpha$ and Pf$\beta$ lines (from HII
regions), H$_2$ v=0-0, S(9), S(10), and S(11) and CO v=1-0 ro-vib lines at
4.665$\mu$m (from shocked molecular gas).  The 4.5$\mu$m band contains no PAH
features and the brightest sources in this band are also stars.  The 5.8$\mu$m
band contains a strong PAH feature at 6.2$\mu$m which can dominate the diffuse
emission except very close to O stars where PAHs are destroyed (see \S 4.1).
Near hot O stars, strong contributions from thermally emitting dust plus a
small contribution from stochastically heated small grains probably dominate
the diffuse emission in the 5.8$\mu$m band.  The 8.0$\mu$m band contains two
very strong PAH features at 7.7 and 8.6$\mu$m which dominate the diffuse
emission in this band, although near hot stars it may be dominated by thermal
dust emission with little or no PAH emission.  Within ionized zones, Hoare
(1990) and Hoare et al. (1991) have shown that trapped Lyman $\alpha$ heated dust can
maintain T $>$ 100 K throughout the ionized region.  Because the emission
process that dominates in each IRAC band depends on the environment, band
ratios provide a powerful tool to measure the extent of each environment.

The main focus of this study is to use wavelength-dependent distributions of
MIR diffuse dust emission and high resolution radio continuum emission to
determine the PAH destruction radius of three bubbles and to trace the
relative sizes and locations of 1) the HII region, 2) the hot thermally
emitting dust, and 3) the location and extent of the PDR region associated
with each bubble.  In section 2 we introduce the different data sets used for
this analysis. In section 3, we discuss results of the observations toward
three bubbles. In section 4, we estimate the PAH destruction radius and dust
temperature. In section 5, we examine the stellar population within each of
the bubbles to identify the ionizing star(s) and any young stellar objects
(YSOs) associated with the bubbles. Conclusions are summarized in section 6.

\section{Data}
Data have been assembled from three imaging surveys of the galactic plane:
GLIMPSE, MIPSGAL, and MAGPIS. The latter two surveys were chosen based on
their resolution, sensitivity, wavelength coverage and overlap with the
GLIMPSE survey which provided the basis for the Churchwell et al. (2006)
bubble catalog. The GLIMPSE survey (Benjamin et al.  2003) imaged the inner
Galactic plane using the IRAC camera (Fazio et al. 2004) on the Spitzer Space
Telescope (Werner et al. 2004). The survey covered 10$^\circ$ $< |l| <$
65$^\circ$ and $|b| <$ 1$^\circ$ at 3.6 $\mu$m, 4.5 $\mu$m, 5.8 $\mu$m and 8.0
$\mu$m with resolution from 1.5'' (3.6$\mu$m) to 1.9'' (8.0$\mu$m). Mosaicked
images were produced using the IPAC Montage program in the GLIMPSE pipeline
after image artifacts were removed. A Point Source Archive was produced of
all point sources detected above the 5$\sigma$ level, about 48 million
sources. See the GLIMPSE Data Products
Description\footnote{www.astro.wisc.edu/sirtf/glimpse1\_dataprod\_v2.0.pdf}
for further details.

MIPSGAL (Carey et al. 2005) is a survey of the inner Galactic plane
(10$^\circ$ $< |l| <$ 65$^\circ$ and $|b| <$ 1$^\circ$) at 24 $\mu$m and 70
$\mu$m using the MIPS (Multiband Imaging Photometer for Spitzer) instrument
(Rieke et al. 2004) on the Spitzer Space Telescope. MIPS has a resolution of
5'' at 24 $\mu$m and 15'' at 70 $\mu$m. We only analyzed the 24 $\mu$m
emission because the 70 $\mu$m appears complex enough to warrant separate
analysis in an upcoming paper.


The Multi-Array Galactic Plane Imaging Survey (MAGPIS; Helfand et al. 2006)
used the VLA in B, C and D configurations combined with the Effelsberg 100m
single dish data to obtain high resolution radio images (with no zero spacing
problems) at 1.4 GHz continuum. The survey covered 5$^\circ$ $< l <$
32$^\circ$ and $|b| <$ 0.8$^\circ$ with a resolution of 6'' and a 1$\sigma$
sensitivity of $\sim$0.3 mJy. Full polarization was preserved.

\section{Results}
We now present the relative distribution of MIR and 20 cm emission for three
bubbles selected because they represent different morphological shapes and/or
show possible evidence for triggered star formation.  They are N10, N21 and
N49 in the Churchwell et al. (2006) bubble catalog.

\subsection{N10}
\begin{figure}
\caption{N10, 24 $\mu$m (red), 8 $\mu$m (green) and 4.5 $\mu$m (blue). 20 cm
  (contours) in bottom panel. Note that the 24 $\mu$m emission is saturated at
  the center of the image. The white dashed line on the top figure indicates
  the location of the cross-cut in figures 2-4.}
\label{n10color}
\plotone{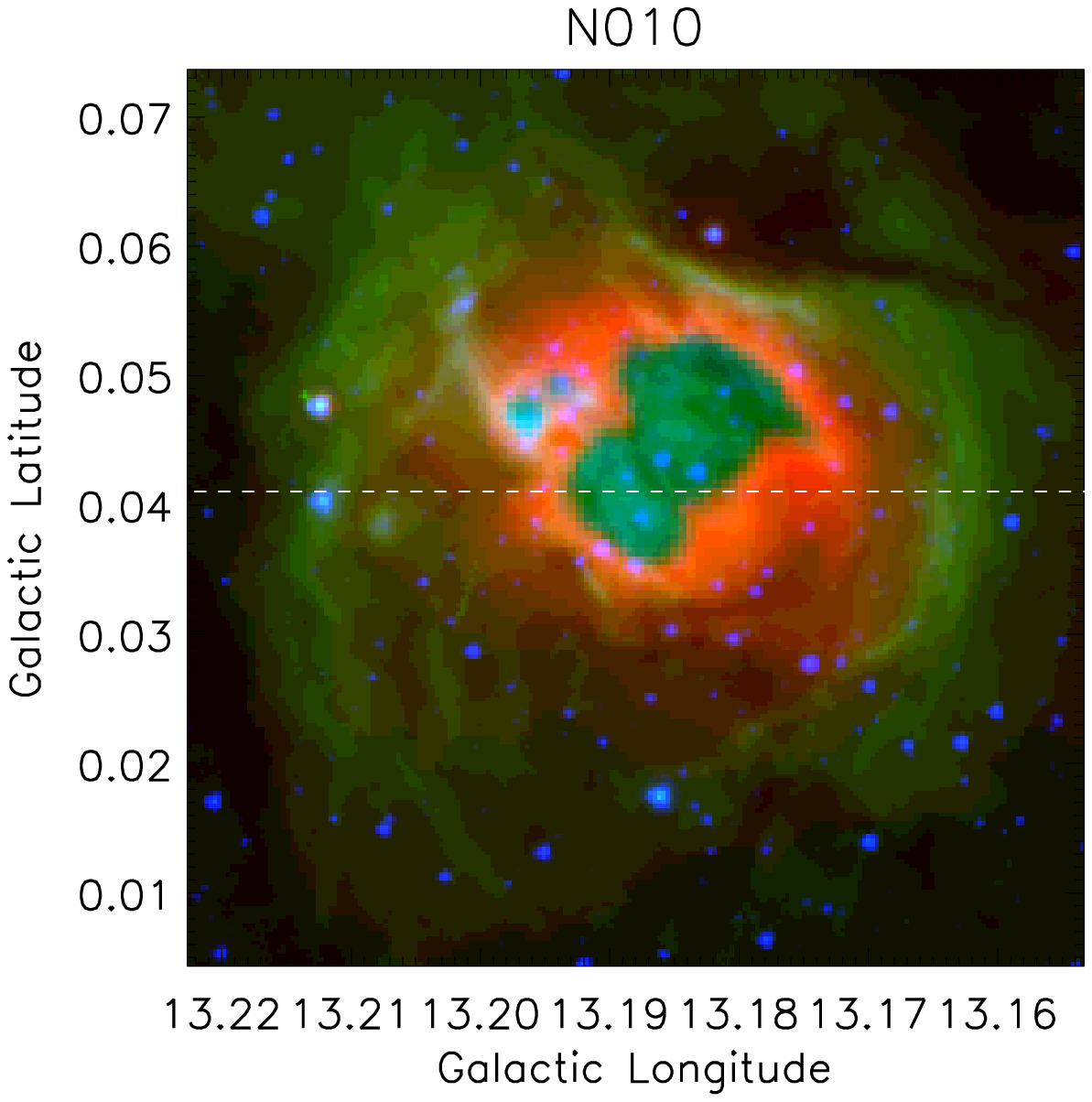}
\plotone{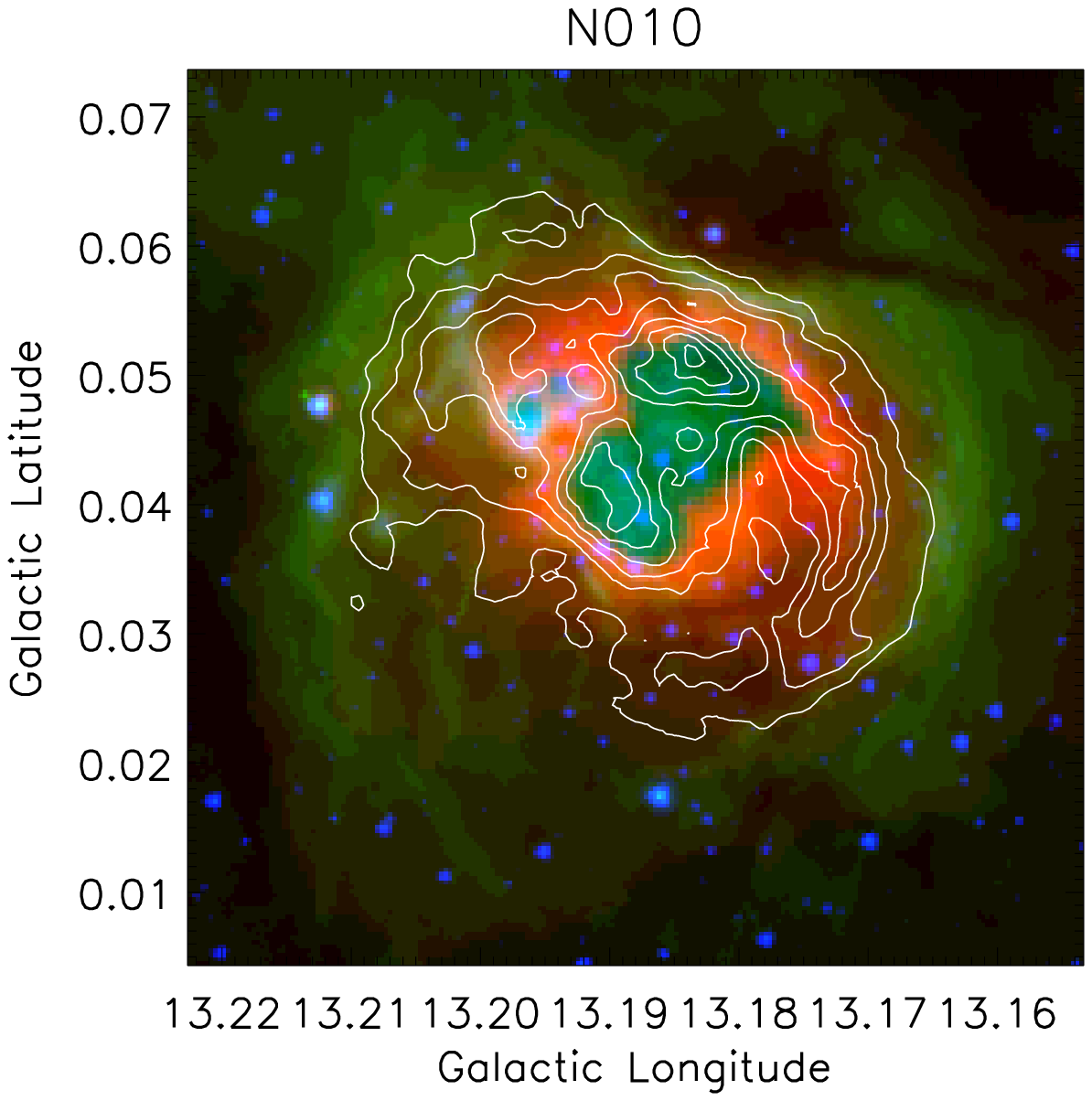}
\end{figure}

N10 is a bright MIR and radio continuum bubble having elliptical or slight
cometary shape with an opening at galactic position angle of 160$^\circ$ (see
Fig.  \ref{n10color}).  Its kinematic distance is 4.9$\pm$0.5 kpc (Churchwell
et al.  2006 and references therein).  At 1.4 GHz it has an integrated flux
density of 7.58 Jy (Helfand et al 2006) and we measure an average angular
radius to the background of 1.26' ($\sim$1.8 pc).  Using the relation of
(Rohlfs \& Wilson, 2006):
\begin{eqnarray*}
\frac{N_{Ly}}{s^{-1}} &=& 6.3 \times 10^{52} \left(\frac{T_e}{10^4}\right)
\left(\frac{\nu}{GHz}\right)^{-0.1} \left(\frac{L_\nu}{10^{20} W
    Hz^{-1}}\right),
\end{eqnarray*}
we calculate 1.6 $\times$10$^{49}$ ionizing photons per second are necessary
to maintain ionization, equivalent to a single O5V star (Martins, Schaerer \&
Hillier 2005, hereafter MSH05). No correction for extinction was made.

\begin{figure}
\caption{N10: Slice at latitude b=0.04$^\circ$. 20 cm (solid, magnified
  10$^6$), 24 $\mu$m (dotted) and 8 $\mu$m (dashed, magnified 5x). The location
  of the 8 $\mu$m shell and central 24 $\mu$m hot dust emission are indicated.
  Note that the 24 $\mu$m emission is saturated in the center of the slice,
  resulting in the strong dip and missing data between longitudes
  13.18$^\circ$ and 13.20$\circ$.}
\label{n10lat1}
\plotone{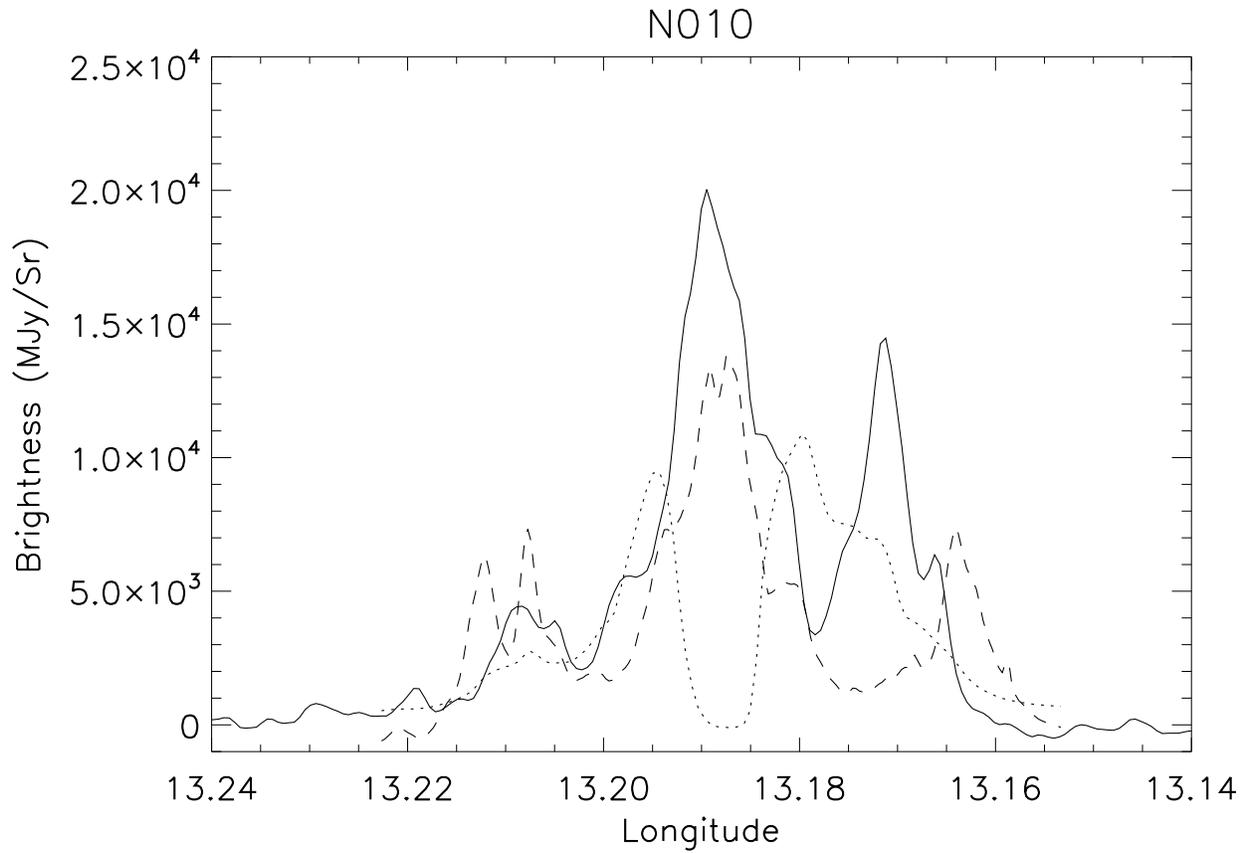}
\end{figure}

\begin{figure}
\caption{N10: Slice at latitude b=0.04$^\circ$. 3.6 $\mu$m (solid), 4.5
  $\mu$m (dotted) and 5.8 $\mu$m (dashed). The spikes in 3.6 $\mu$m and 4.5
  $\mu$m emission indicate stars.}
\label{n10lat2}
\plotone{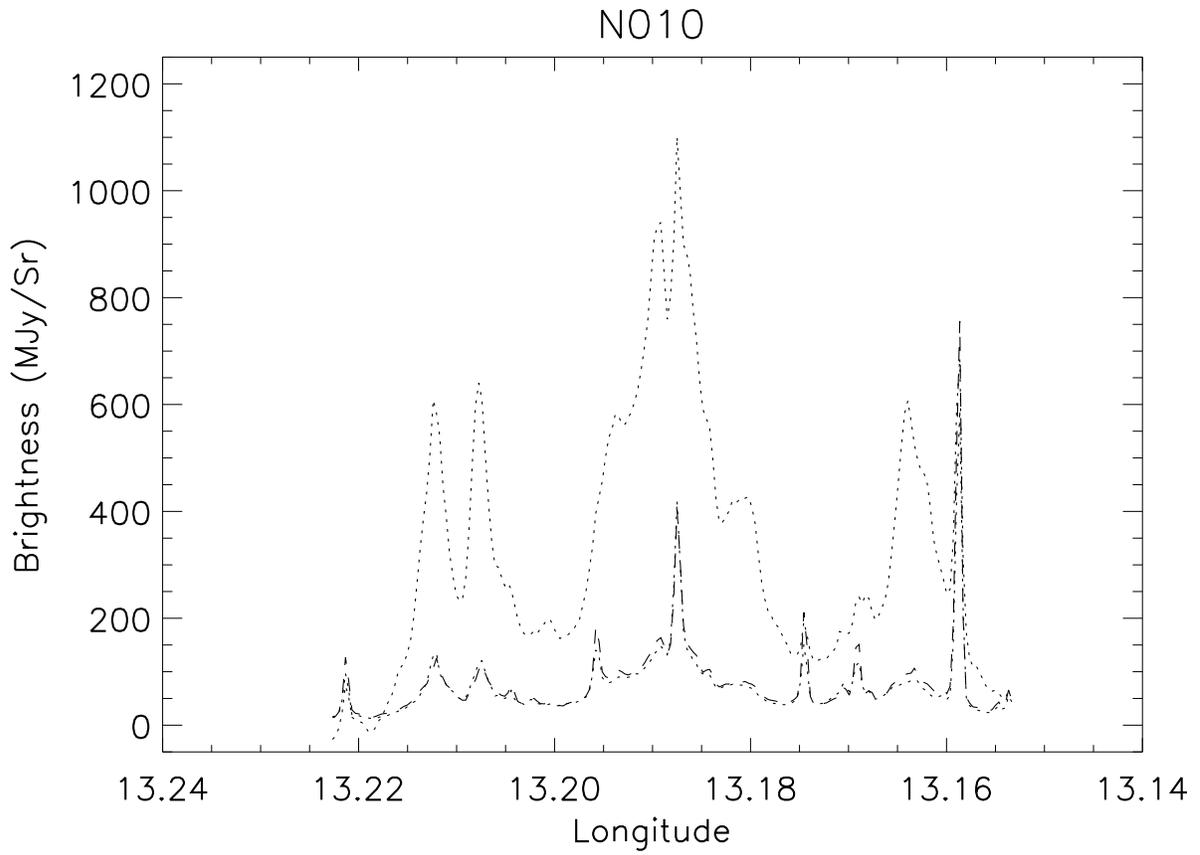}
\end{figure}

The relative distributions of emission as a function of wavelength are
illustrated in Figs. \ref{n10lat1}-\ref{n10lat2}.  The bubble is surrounded by
an 8$\mu$m emission shell with angular radius (out to the Galactic background
level) of 1.8' (2.6 pc).  The radius to the inner face of the 8 $\mu$m shell
is 1.2' (1.7 pc). The 8$\mu$m emission rises very sharply on the inner face of
the shell and declines gradually with increasing radius.  We postulate that
this ring is dominated by PAH emission and the inner face of the shell defines
the destruction radius of PAHs (see \S 4.1).  It is bright because swept-up
interstellar dust densities are high here and the dust is exposed to a large
flux of soft UV radiation (non-H ionizing photons) that excites PAHs but is
not energetic enough to destroy PAHs.  The slow fall-off beyond the inner face
of the 8$\mu$m shell represents the PDR region of the bubble, primarily
delineated by PAH emission.

Inside the 8$\mu$m shell, 24$\mu$m, 20 cm, 8.0 $\mu$m and 5.8 $\mu$m emission
all peak at the same position, showing that hot dust is present inside the HII
region.  Note that the 24 $\mu$m emission is saturated at the center. Clearly
the stellar wind(s) have not yet succeeded in clearing out or destroying all
the dust in the bubble.  There is also some diffuse 5.8 $\mu$m and 8 $\mu$m
emission inside the bubble.
Figure \ref{n10lat2} shows that the 3.6 $\mu$m and 4.5$\mu$m emission vary
together because both are dominated by stellar emission.  The 5.8 $\mu$m and
8.0$\mu$m band vary together also, although 5.8$\mu$m emission is generally
fainter than 8.0$\mu$m emission. The relative distribution of emission at
different wavelengths in N10 (i.e. an 8$\mu$m shell enclosing 24$\mu$m and
radio continuum emission) is a general property of all the bubbles for which
we have MAGPIS and MIPSGAL data.

\subsection{N21}
\begin{figure}
\caption{N21, 24 $\mu$m (red), 8 $\mu$m (green), 4.5 $\mu$m (blue) and 20 cm
  (contours) in the bottom panel.The white dashed line on the top figure
  indicates the location of the cross-cut in figures 6 \& 7.}
\label{n21color}
\plotone{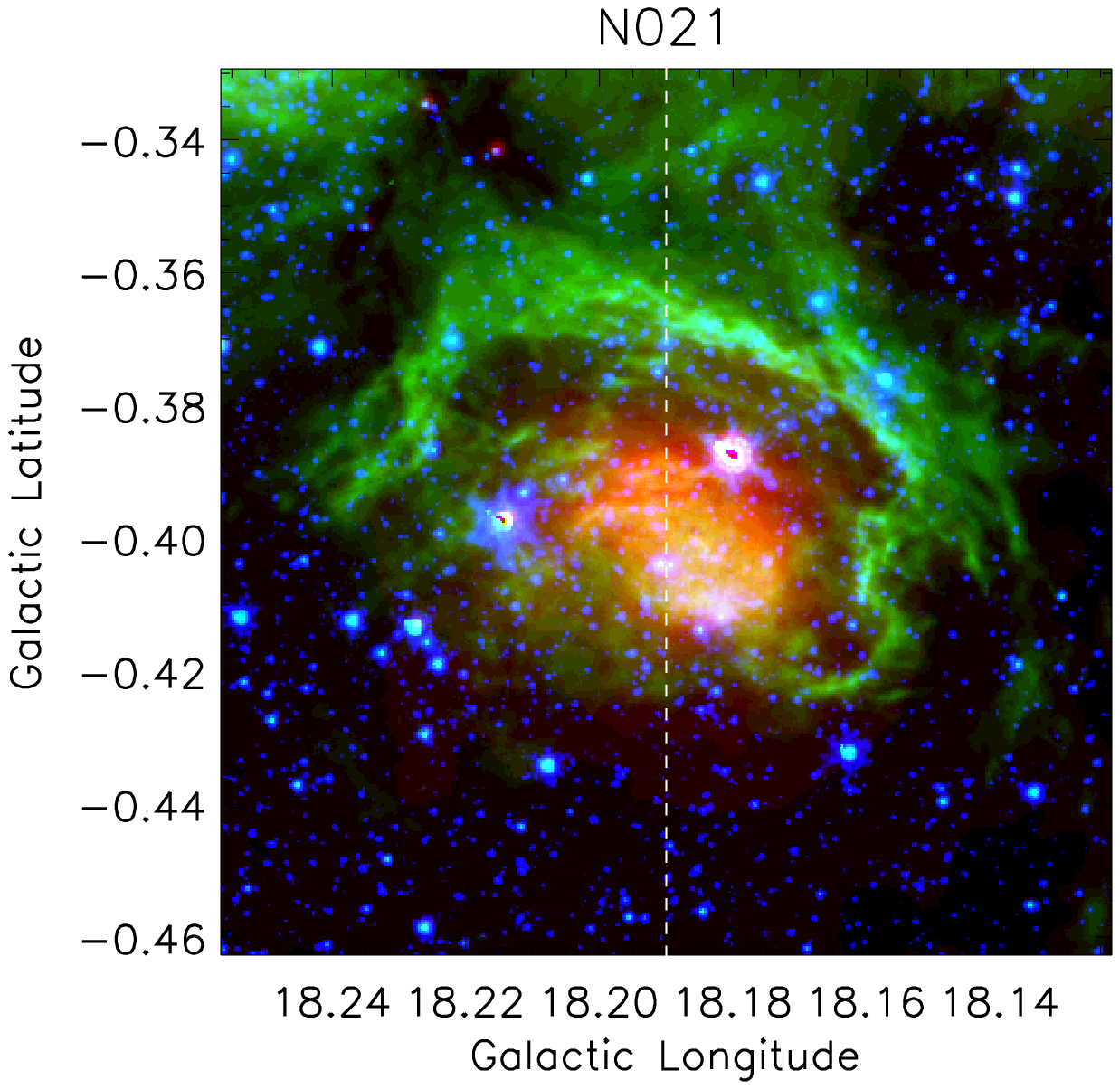}
\plotone{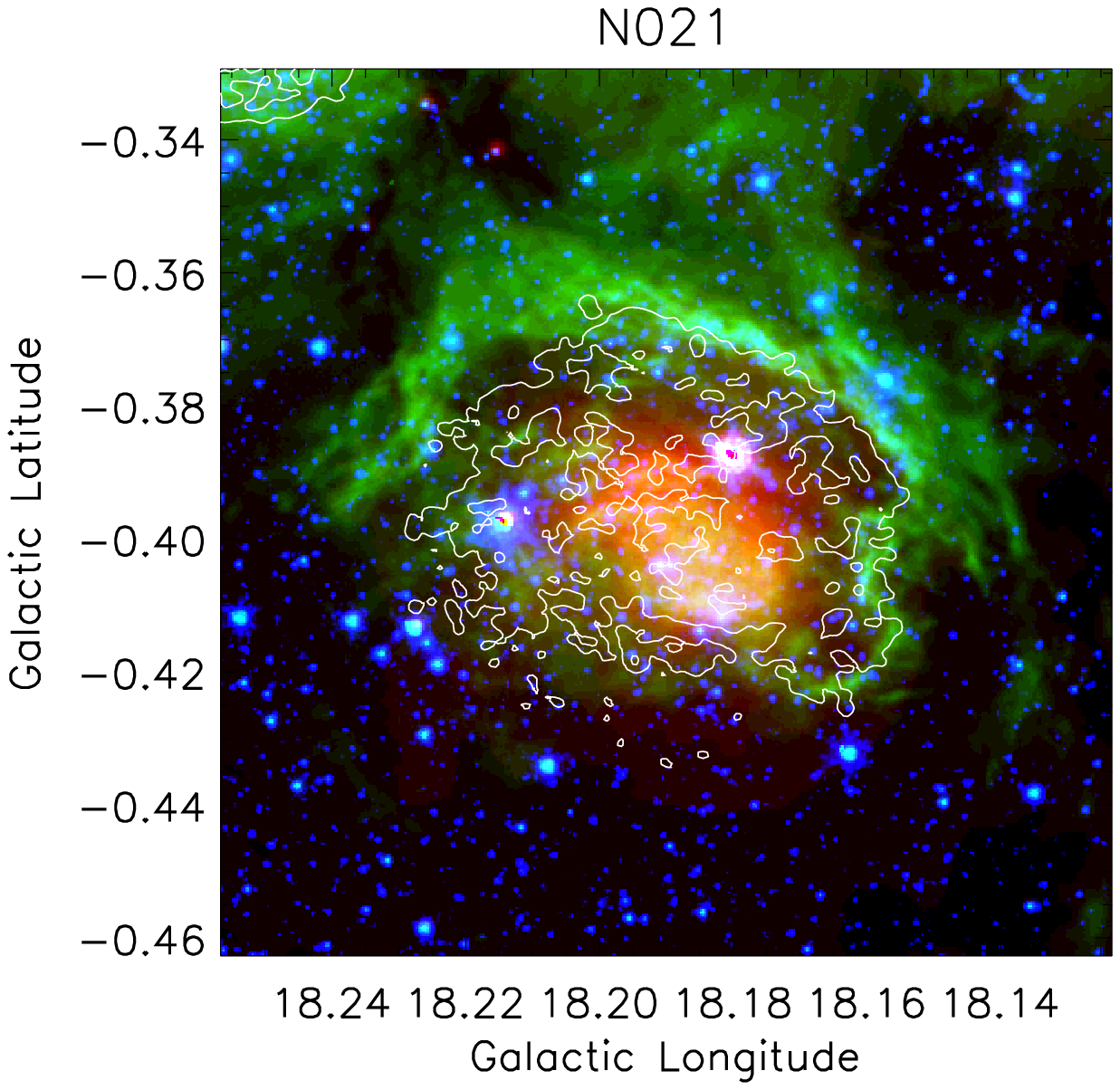}
\end{figure}

N21 has a cometary morphology in 8 $\mu$m emission (see Fig \ref{n21color}),
but otherwise has similar relative spatial distributions with wavelength as
N10 with the exception that N21 is not bounded at 8$\mu$m along the lower half
of the bubble. N21 has a kinematic distance of 3.7$\pm$0.6 kpc (Churchwell et
al.  2006). At 1.4 GHz it has an integrated flux density of 7.2 Jy and an
angular radius of 2.6' (2.8 pc) (Helfand et al.  2006). It requires
8.5$\times$10$^{48}$ ionizing photons s$^{-1}$ to maintain ionization,
equivalent to a single O6V star (MSH05).

\begin{figure}
\caption{N21: Slice at longitude l=18.19$^\circ$. 20 cm (solid, magnified
  10$^6$x), 24 $\mu$m (dotted) and 8 $\mu$m (dashed, magnified 5x)}
\label{n21lon1}
\plotone{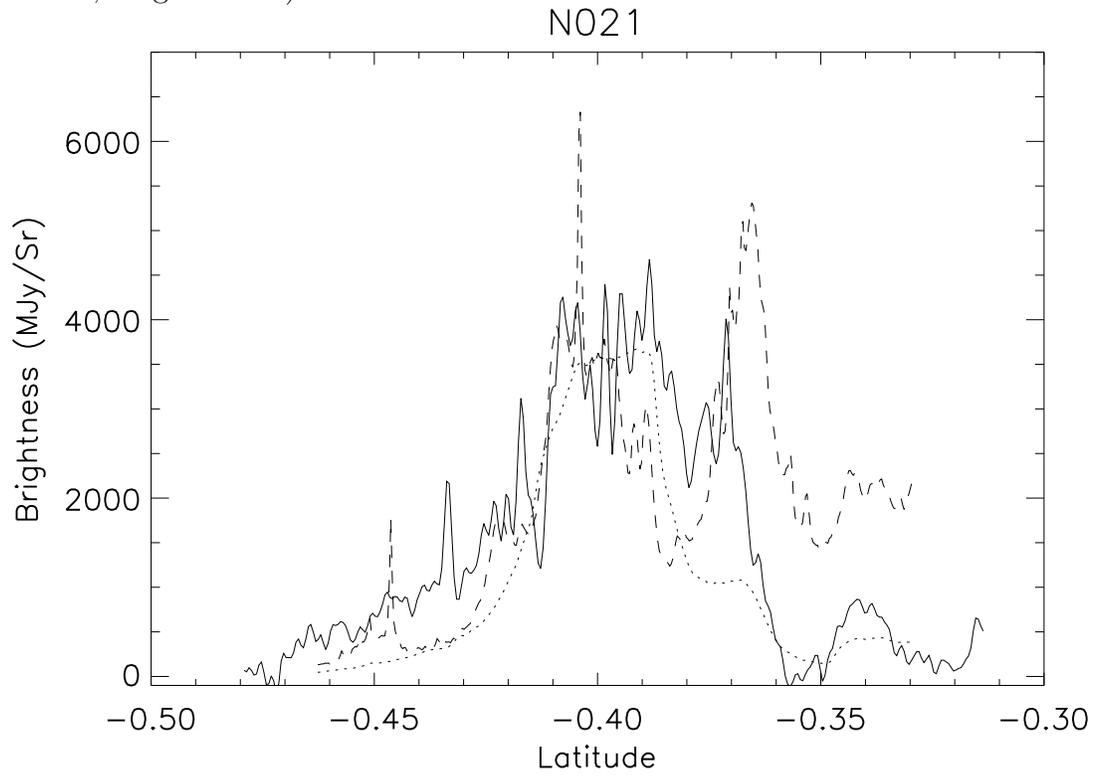}
\end{figure}

\begin{figure}
\caption{N21: Slice at longitude l=18.19$^\circ$. 3.6 $\mu$m (solid), 4.5
  $\mu$m (dotted) and 5.8 $\mu$m (dashed). The spikes in 3.6 $\mu$m and 4.5
  $\mu$m emission indicate stars.}
\label{n21lon2}
\plotone{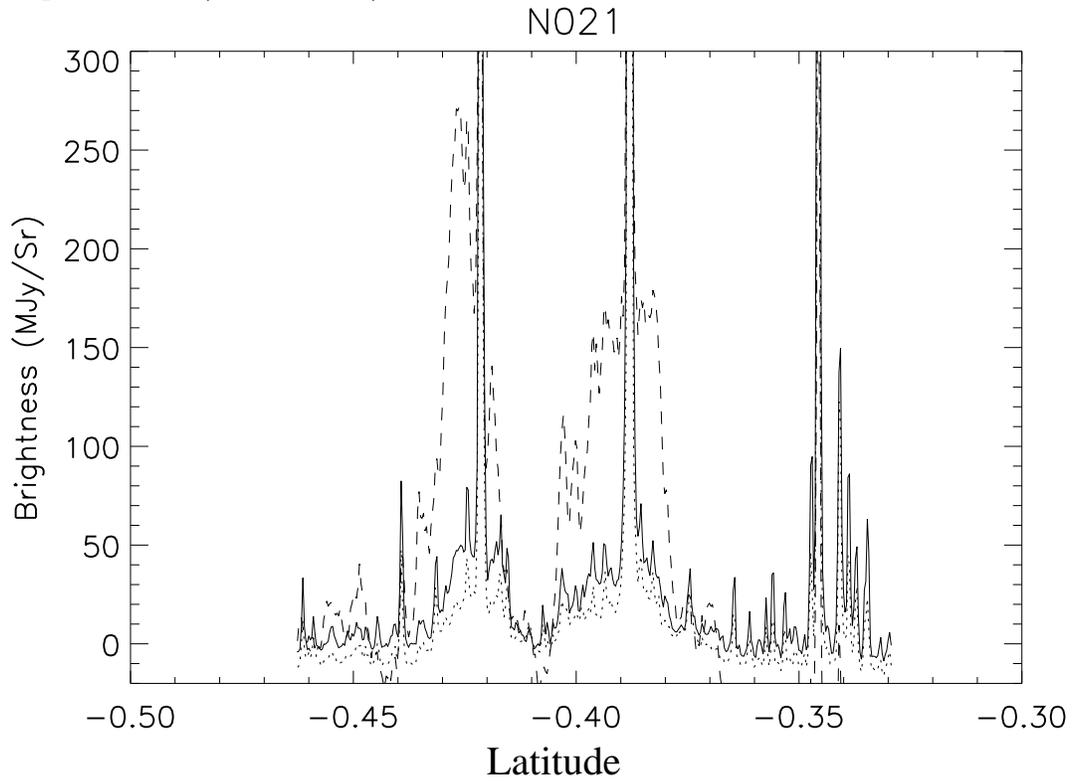}
\end{figure}

The 8 $\mu$m shell has a radius to the inner face of 2.1' (2.2 pc) and a
radius out to the background level of of 3.0' (3.2 pc). The clearest
perspective on the relative emission distributions with wavelength are the
slices at constant longitude (see Fig.  \ref{n21lon1}-\ref{n21lon2}). Figure
\ref{n21lon1} shows that 8$\mu$m emission is located above the 20 cm and 24
$\mu$m emission. 20 cm and 24$\mu$m emission increase and decrease in near
unison. The 5.8 $\mu$m and 8.0 $\mu$m emission also vary together (see
Figures~\ref{n21lon1} \& \ref{n21lon2}). 5.8 $\mu$m and 8.0 $\mu$m are present
both inside the bubble and along the top boundary.  Inside the bubble, the 8
$\mu$m flux is likely dominated by both PAH emission from the front-side and
backside of the bubble and by hot dust grains inside the bubble.  We measure
the integrated flux density at 8$\mu$m inside the 8 $\mu$m shell (0.395$^\circ
< b <$ 0.41$^\circ$) to be 228 Jy and the integrated flux density at 24$\mu$m
in the same region to be 599 Jy. The former value may be over-estimated by
$\sim$30\% because of IRAC diffuse calibration errors (Cohen et al. 2007). The
dust temperature cannot be determined because of PAH contamination at 8
$\mu$m.
  
The relative distributions of hot dust, excited PAHs and the PDR region in N21
is similar to N10. The key difference between N10 and N21 is the absence of 8
$\mu$m emission along the southern half of the bubble. The detected 24 $\mu$m
and 20 cm emission along the southern interior, however, implies the existence
of ionizing photons.  The near absence of the 8 $\mu$m shell along the lower
half of this bubble thus implies the absence of PAHs.  This may be because the
ambient ISM density in this direction is low enough that the PAHs have either
been destroyed by direct exposure to hard stellar UV radiation and/or blown
out from the bubble far enough that the PAHs cannot be excited by the stellar
radiation due to geometric dilution.
  

\subsection{N49}
N49 is a bright MIR bubble surrounding a radio HII region that has an almost
spherical structure.  It has a kinematic distance of 5.7$\pm$0.6 kpc
(Churchwell et al.  2006). At 1.4 GHz it has an integrated flux density of 2.8
Jy and an angular radius out to the background of 1.5' ($\sim$2.5 pc) (Helfand
et al. 2006). 7.8 $\times$10$^{48}$ ionizing photons s$^{-1}$ are necessary to
maintain its ionization, equivalent to a single O6V star (MSH05). The radius
to the inner face of the 8 $\mu$m shell is 1.2' (2.0 pc) and out to the
background level is 1.7' (2.3 pc).

\begin{figure}
\caption{N49, 24 $\mu$m (red), 8 $\mu$m (green), 4.5 $\mu$m (blue) and 20 cm
  (contours) in bottom panel. The white dashed line on the top figure
  indicates the location of the cross-cut in figures 10 \& 11.}
\label{n49color}
\plotone{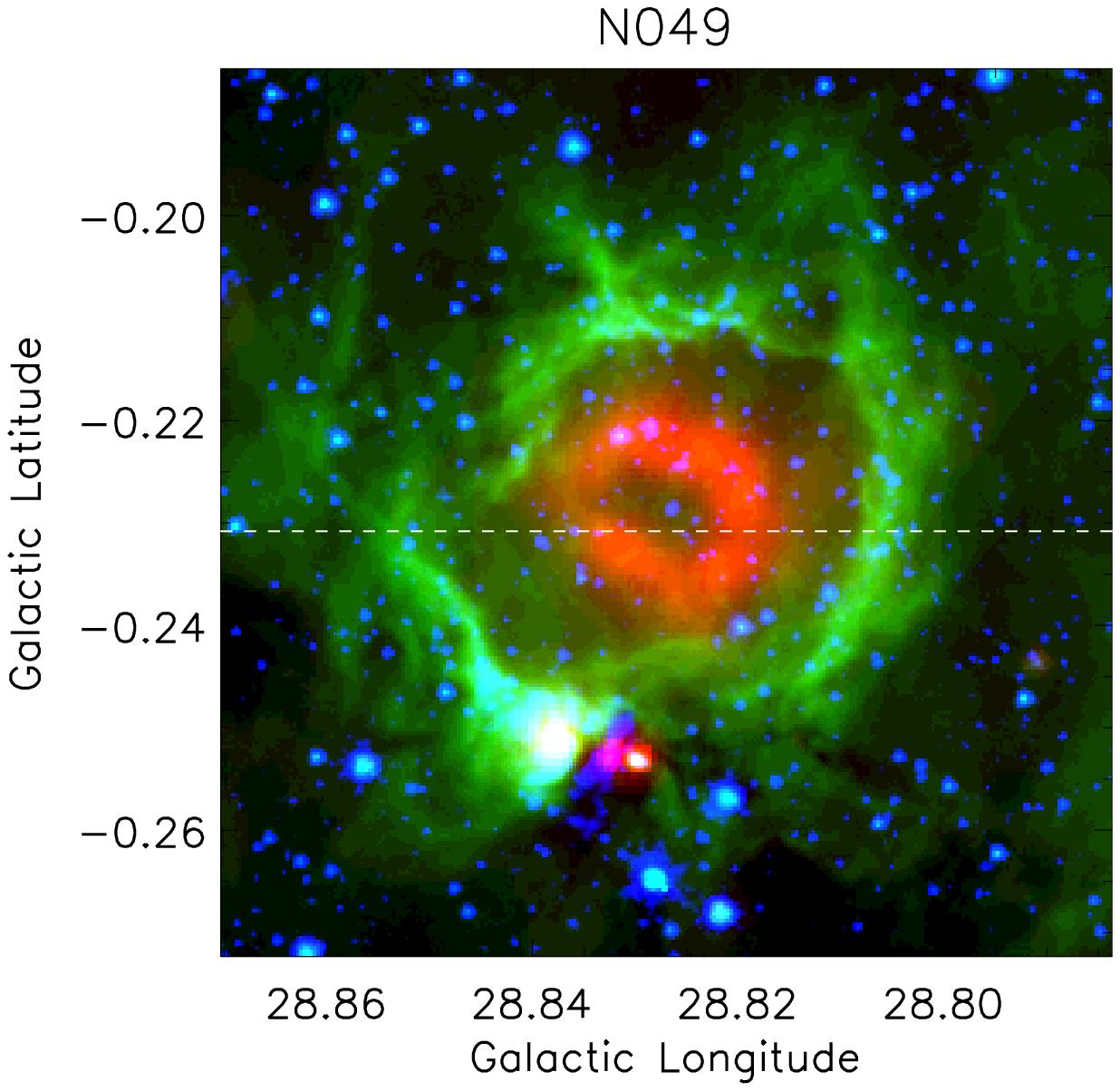}
\plotone{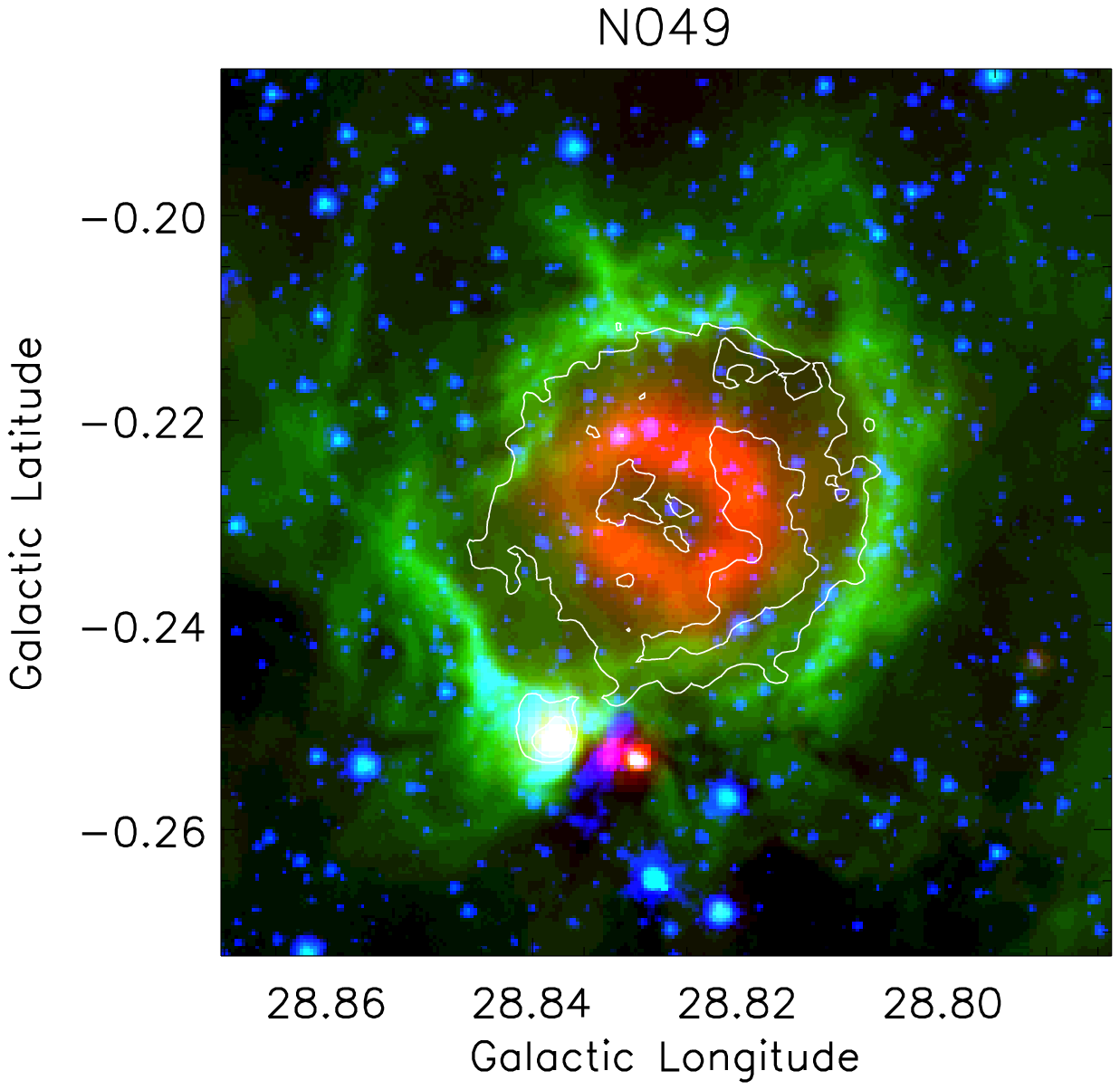}
\end{figure}

\begin{figure}
\caption{N49: Slice at latitude b=-0.23$^\circ$. 20 cm (solid, magnified
  10$^6$x), 24 $\mu$m (dotted) and 8 $\mu$m (dashed, magnified 5x). Note that
  there is no central peak at 24 $\mu$m as there is in N10 and N21.}
\label{n49lat1}
\plotone{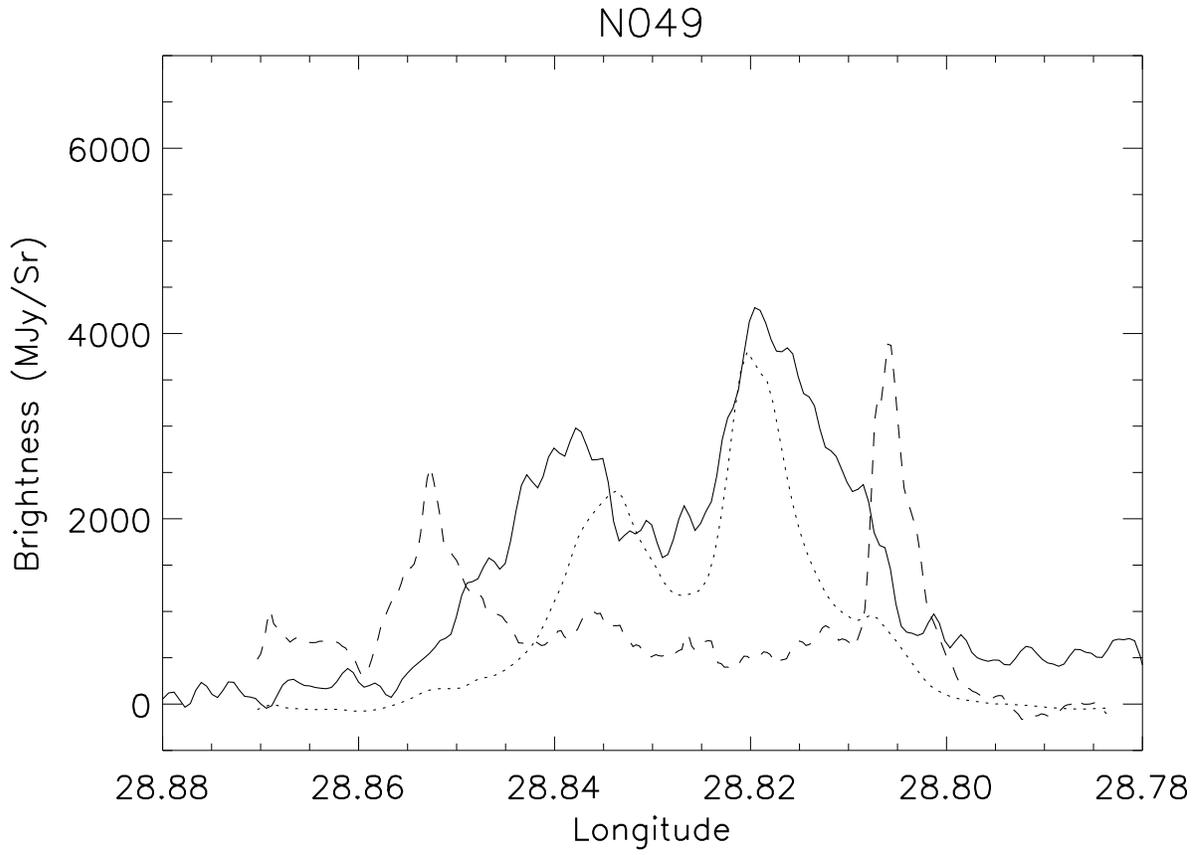}
\end{figure}

N49 has a double-shell structure, the outer traced by 8$\mu$m emission and the
inner traced by 24$\mu$m and 20 cm emission (see Fig \ref{n49color}). As in
N10 and N21, 8$\mu$m emission encloses both the 24$\mu$m and 20 cm emission.
The transition between the 8$\mu$m emission ring and the 24$\mu$m and 20 cm
emission ring can be clearly seen in the slice at constant latitude in Fig
\ref{n49lat1}. The 20 cm and 24$\mu$m emission are coincident, both of which
have a central cavity.  The 24 $\mu$m/20 cm dip appears to be the central
wind-evacuated cavity expected around early-O stars.


\section{Analysis}
We propose the following picture for the IR bubbles: ionized gas with a hot
dust component is surrounded by a PDR containing swept-up interstellar gas,
PAHs, and dust.  The ionized gas is traced by 20 cm free-free emission, the
hot dust within the HII region is bright at 24$\mu$m via thermal continuum
emission.  The IR bubbles are enclosed by a shell of 8$\mu$m emission
dominated by PAH emission features in IRAC bands 3.6, 5,8, and 8.0$\mu$m. The
inner face of the 8$\mu$m shell defines the PAH destruction radius from the
central ionizing star(s). In the following sections, we determine the PAH
destruction radii and PDR shell thicknesses based on $\frac{5.8 \mu m}{4.5 \mu
  m}$ and $\frac{8.0 \mu m}{4.5 \mu m}$ flux density ratios.

 

\subsection{PAH Destruction}

Povich et al. (2007) argued that ratios of IRAC bands that contain strong PAH
emission features (8.0$\mu$m and 5.8$\mu$m bands) to the 4.5$\mu$m band (which
contains no PAH feature) can be used to determine the PAH destruction radius
and define the extent of PDRs around hot stars.  This technique was applied by
Povich et al. (2007) to derive both the PAH destruction region in M17 and the
extent of its PDR because the 8.0 $\mu$m, 5.8 $\mu$m and 3.6 $\mu$m IRAC bands
all contain PAH bands, whereas band-ratios involving the 4.5 $\mu$m PAH-free
band should be especially sensitive to regions containing PAHs.  They
supported their interpretation of these ratios by showing that the
5.8$\mu$m/3.6$\mu$m ratio does not delineate the PDR boundaries. They also
presented IRS spectra that proved the disappearance of PAH features within the
M17 HII region. Povich et al. (2007) were unable to use the 8.0$\mu$m images
of M17 because the detector was saturated over large regions.  We have applied
this technique to N10, N21, and N49. The quantitative ratios are different
from those toward M17 because M17 is a much more luminous region but the
principle is the same. Since N10, N21 and N49 do not saturate the 8.0$\mu$m
detector, we are able to use this band in our analysis as well.

\begin{figure}
\caption{N10: PAH destruction as measured from 5.8 $\mu$m / 4.5 $\mu$m
  (upper-left) and 8.0 $\mu$m / 4.5 $\mu$m (lower-left). The color scale
  ranges from 5 to 12 (upper-left, blue-to-red) and 24 to 44 (lower-left,
  blue-to-red). The contours represent ratios of 8.5 (upper-left) and 25
  (lower-left). Dashed white lines at left indicate position of slices show at
  right.}
\label{n10pahc}
\epsscale{1.2}
\plottwo{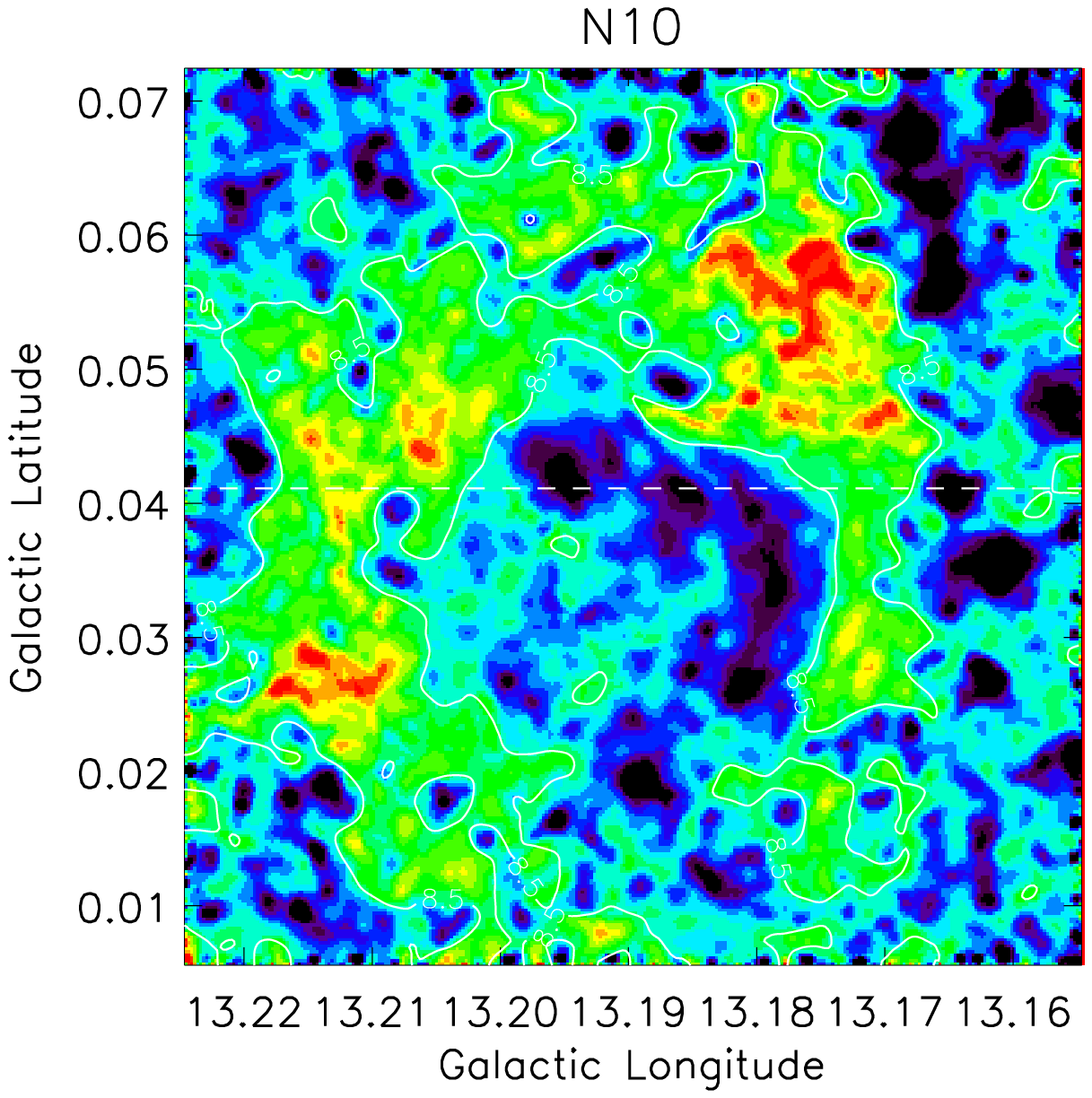}{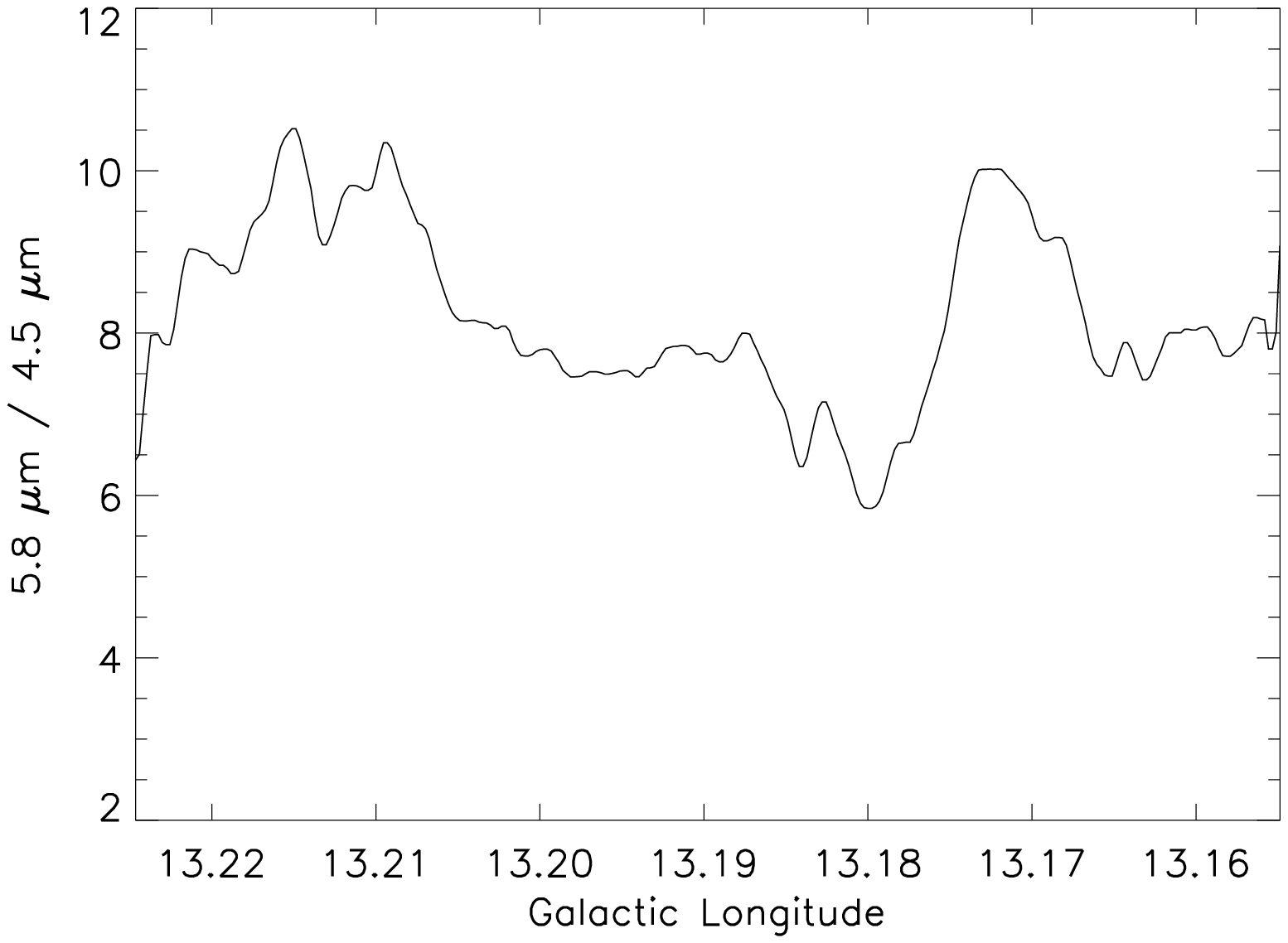}
\plottwo{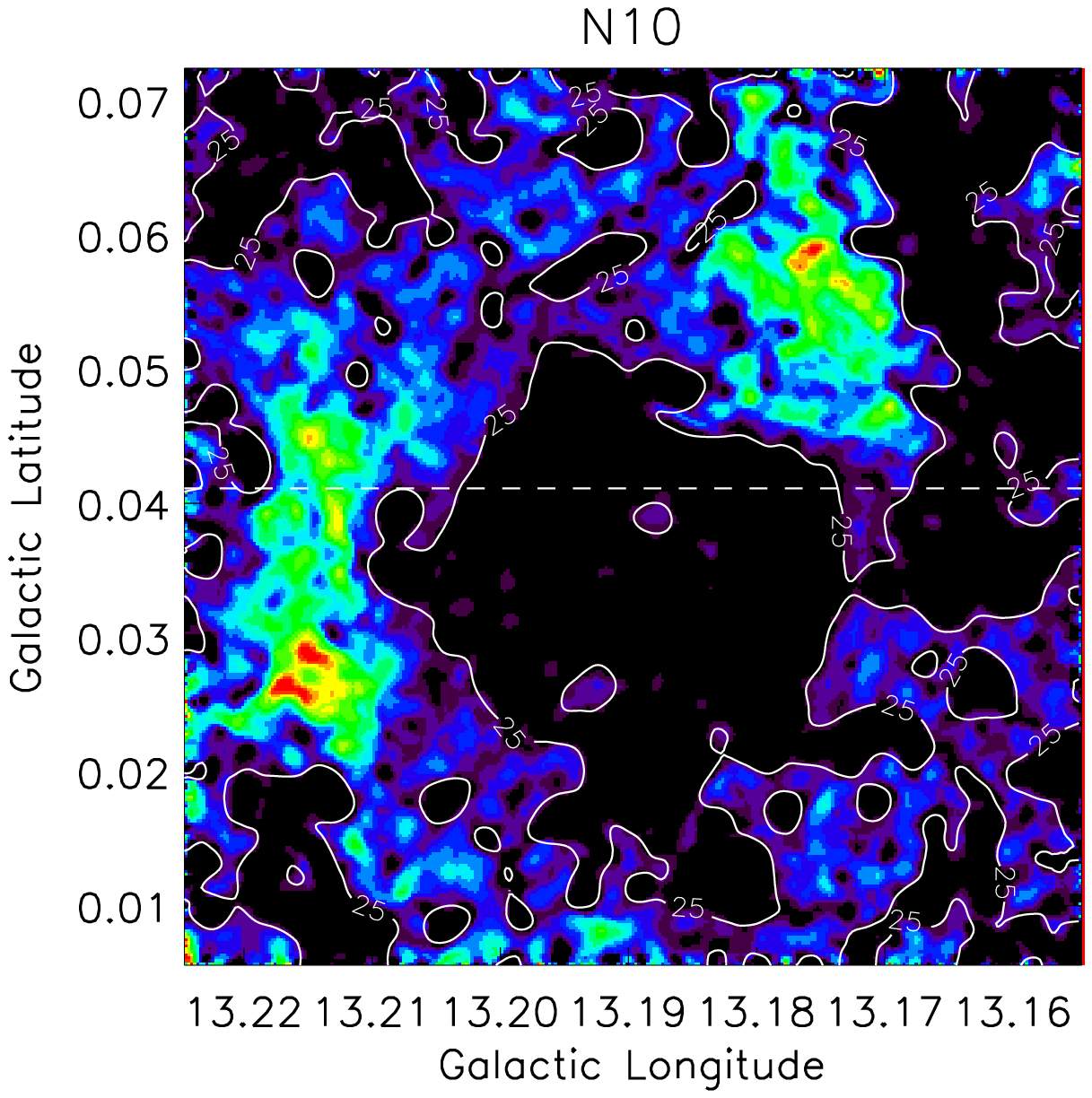}{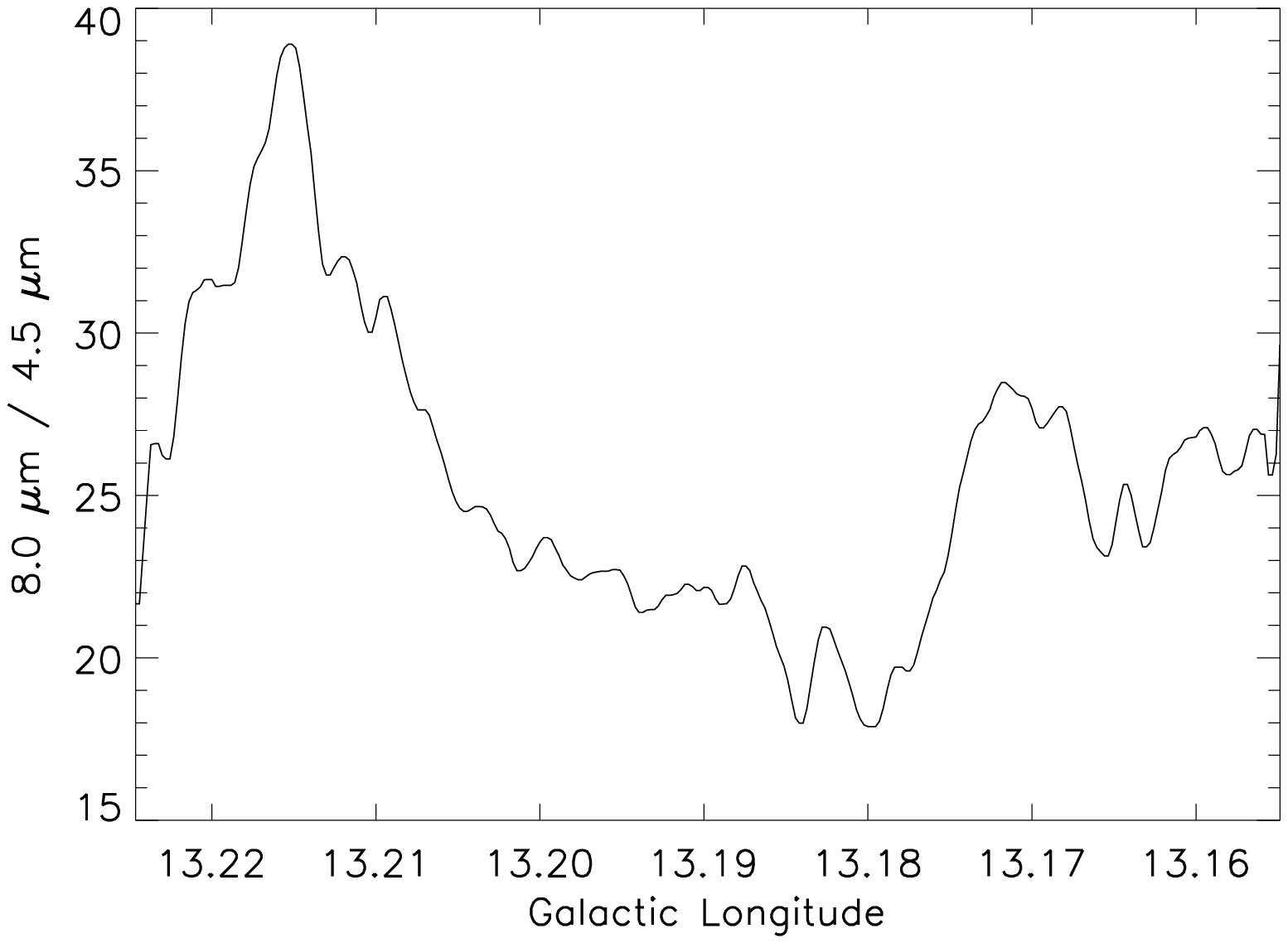}
\end{figure}

\begin{figure}
\caption{N21: PAH destruction as measured from 5.8 $\mu$m / 4.5 $\mu$m
  (upper-left) and 8.0 $\mu$m / 4.5 $\mu$m (lower-left). The color scale
  ranges from 4 to 12 (upper-left, blue-to-red) and 15 to 40 (lower-left,
  blue-to-red). The contours represent ratios of 7.5 (upper-left) and 25
  (lower-left). Dashed white lines at left indicate position of slices show at
  right.}
\label{n21pah}
\plottwo{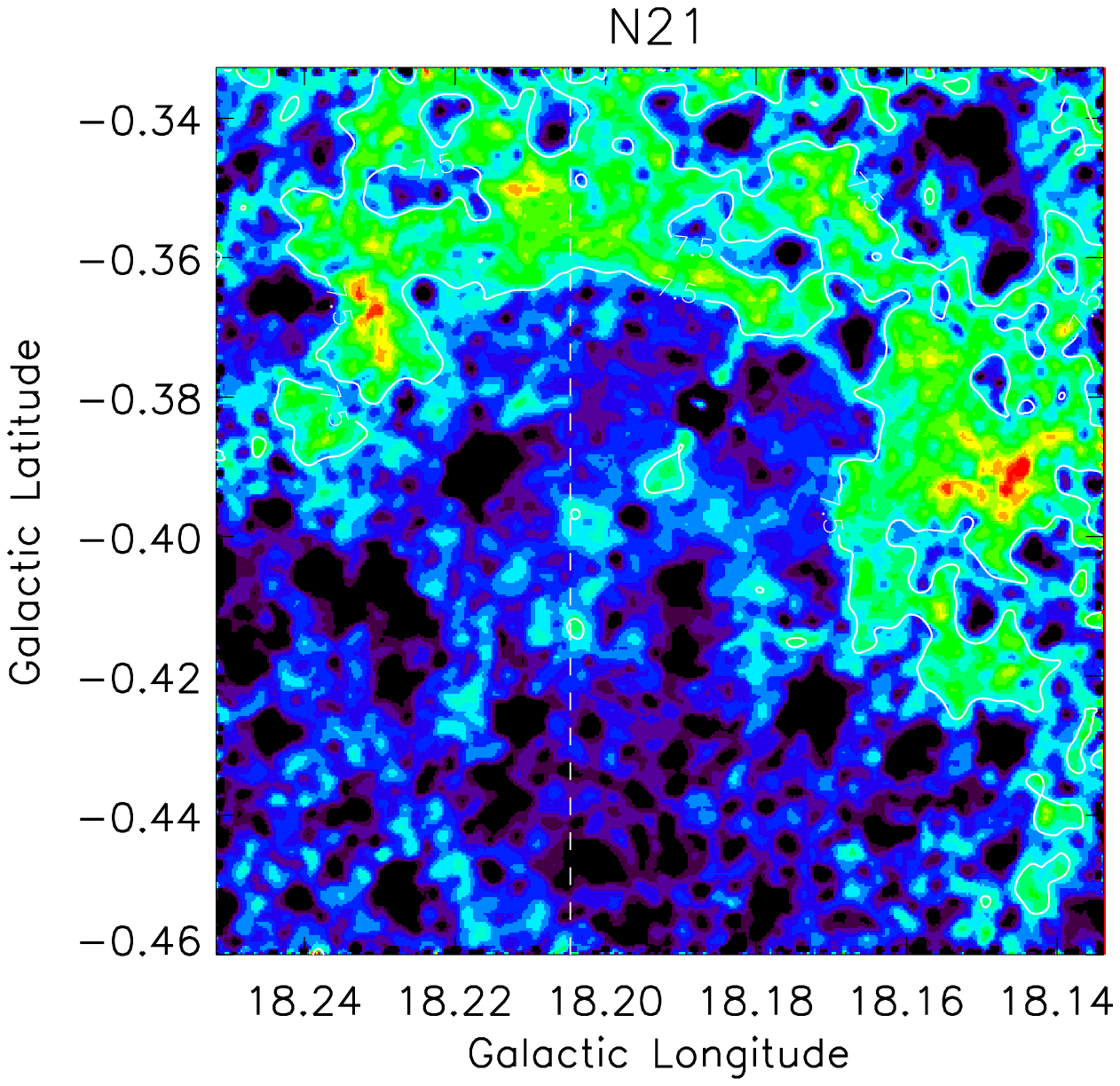}{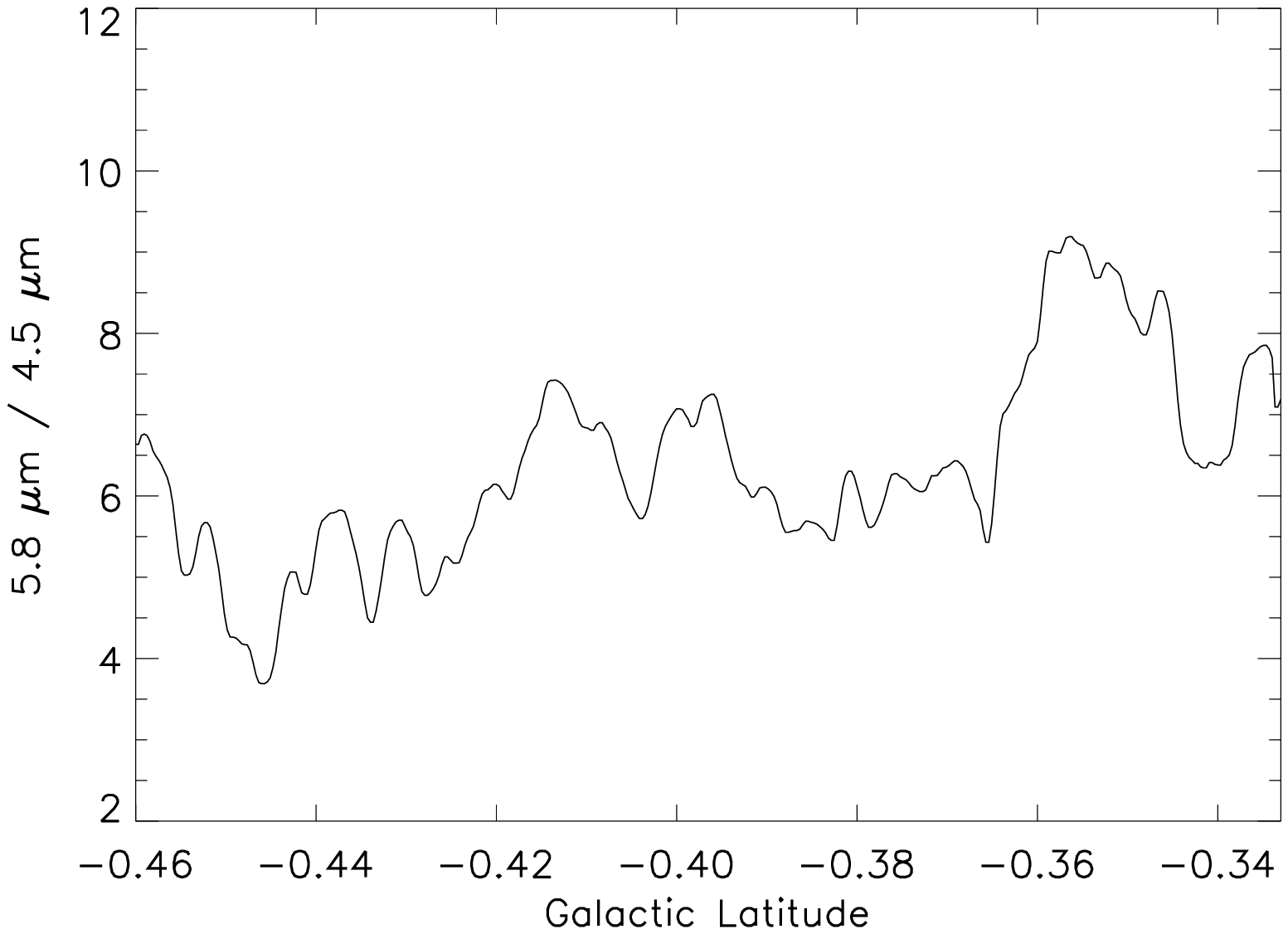}
\plottwo{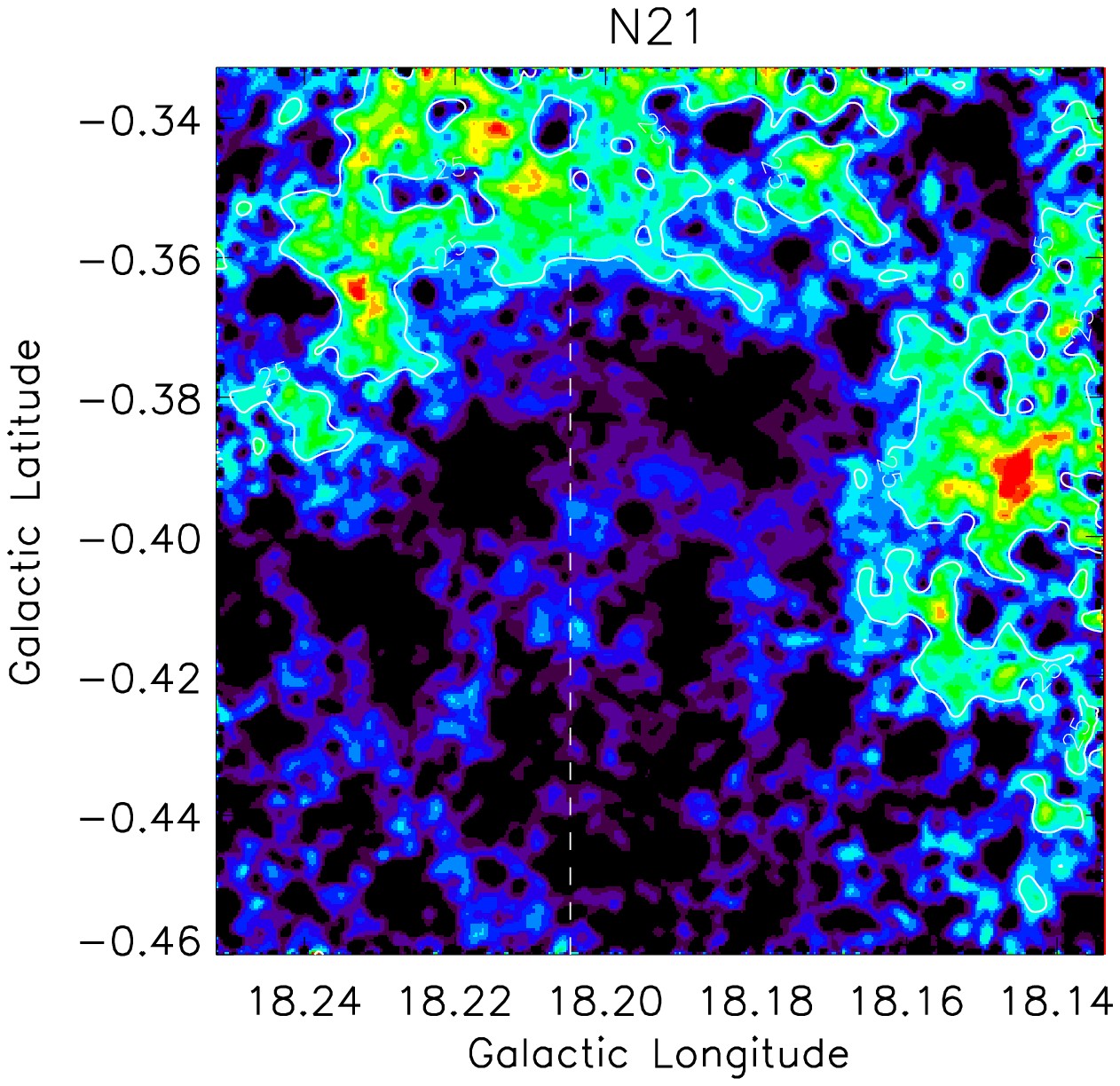}{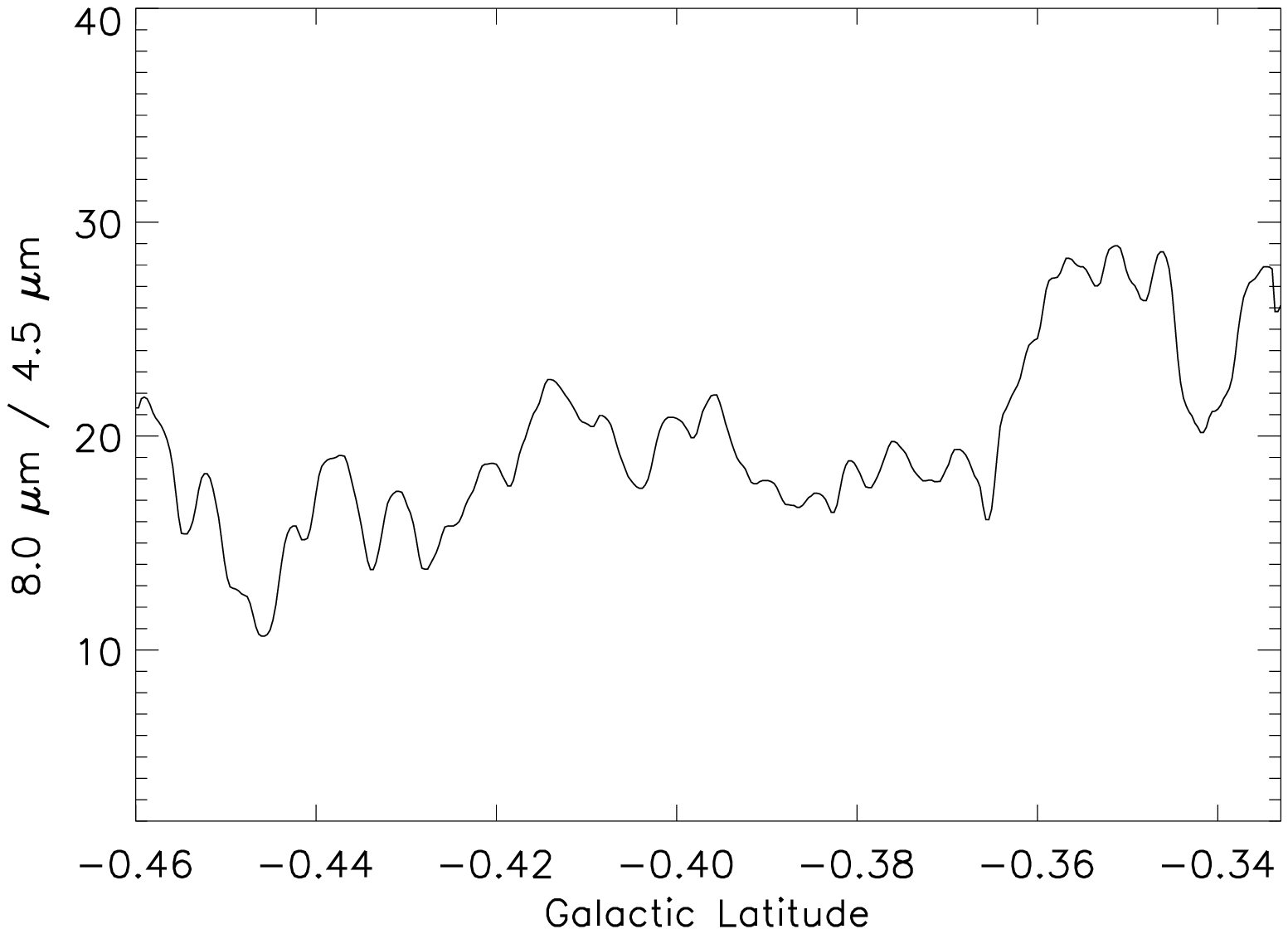}
\end{figure}

\begin{figure}
\caption{N49: PAH destruction as measured from 5.8 $\mu$m / 4.5 $\mu$m
  (upper-left) and 8.0 $\mu$m / 4.5 $\mu$m (lower-left). The color scale
  ranges from 7 to 13 (upper-left, blue-to-red) and 15 to 42 (lower-left,
  blue-to-red). The contours represent ratios of 8.5 (upper-left) and 28
  (lower-left). Longitude slices at constant latitude (upper-right and
  lower-right). Dashed white lines at left indicate position of slices show at
  right.}
\label{n49pah}
\plottwo{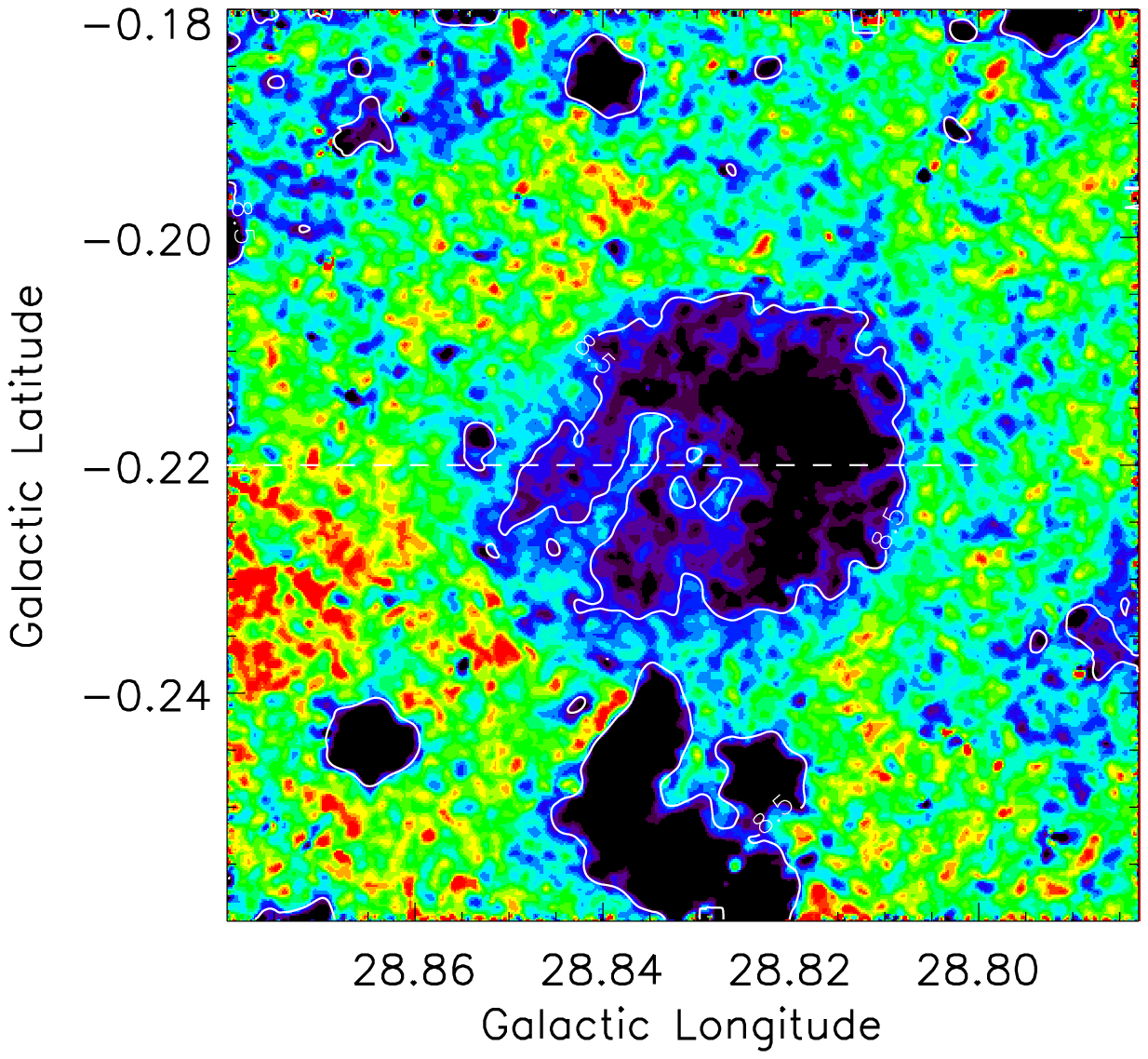}{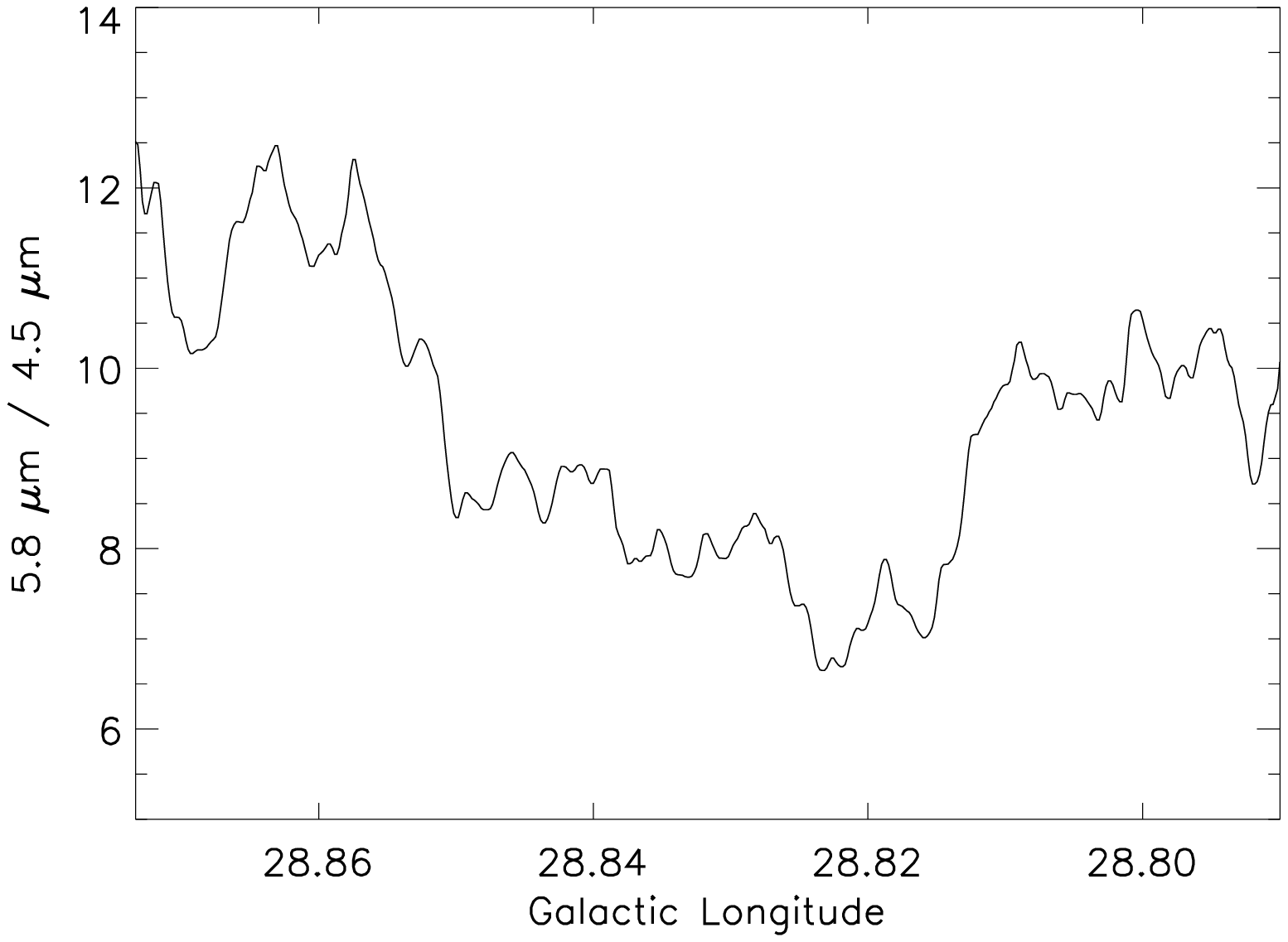}
\plottwo{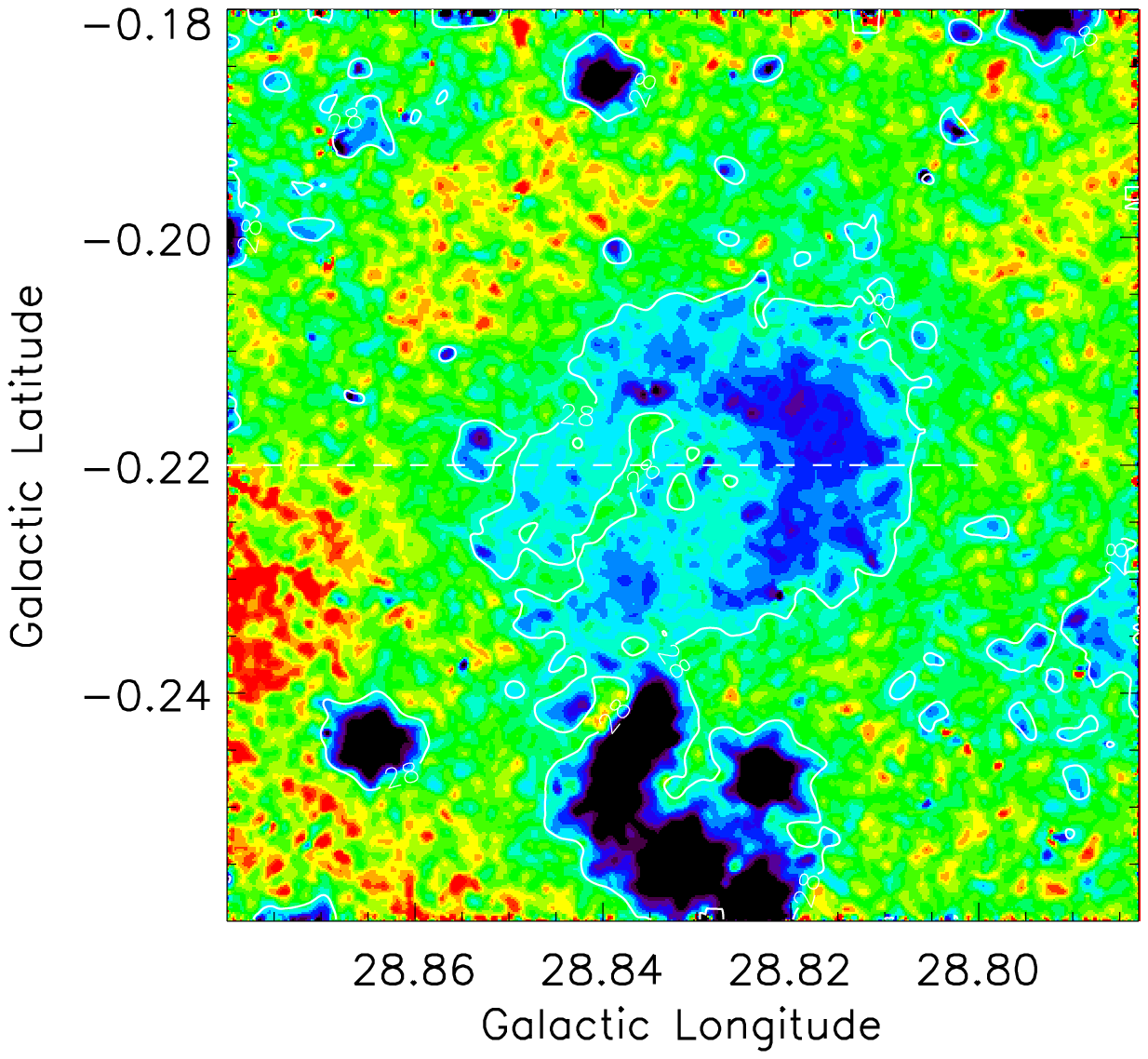}{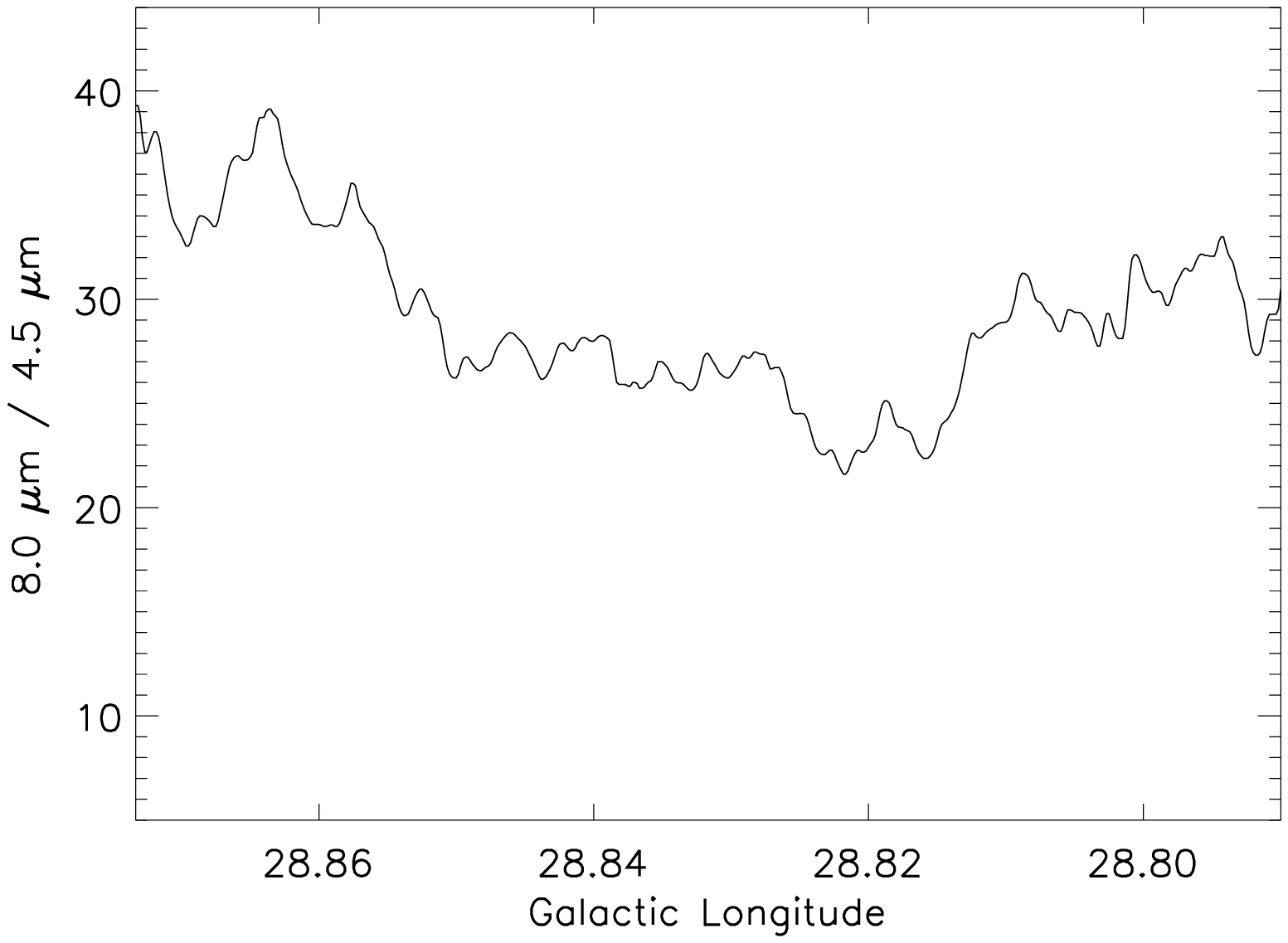}
\end{figure}

Figs. \ref{n10pahc} - \ref{n49pah} show false color images of the 5.8
$\mu$m/4.5 $\mu$m and 8.0 $\mu$m/4.5 $\mu$m band ratios, with accompanying
longitude or latitude cuts (averaged over 20 pixels) for all three bubbles. We
contour the false-color images to indicate the average values that define the
PDR regions.  For each source, the 5.8 $\mu$m/4.5 $\mu$m ratio is 6-7 in the
bubble interiors, while in the inner edge of the 8$\mu$m shells (brightest
part of the PDR region) it is 9-10. The 8.0 $\mu$m/4.5 $\mu$m ratio has a
slightly larger contrast between the shell and the interior, typically 30-35
versus 23-25, respectively. Using the 5.8 $\mu$m/4.5 $\mu$m ratio to determine
the PAH-free region, the bubble interiors have radii of 1.2', 2.2', and 1.5'
for N10, N21 and N49, respectively.  These are about the same as the radii of
the HII regions associated with the bubbles (1.26', 2.6', and 1.5',
respectively).  The typical radial thickness of the PDR regions are 0.6',
0.6', and 0.3', respectively. Roger \& Dewdney (1992) use computer models to
show how the ratio of the PDR outer radius to the HII radius is related to the
ambient density the bubble is expanding into. Based on the ratios 1.4, 1.1 and
1.2 for N10, N21 and N49, we conclude that ambient density is $\sim$10$^3$
cm$^{-3}$. These estimates should be taken with caution, however, since they
are not consistent with the estimated T$_{eff}$ of the ionizing star that we
determine below (see \S 5).  Lastly, there are also some very conspicuous low
ratios (dark spots) in the false color images, especially in the
8$\mu$m/4.5$\mu$m images.  These are locations of bright stars, which are
bright at 4.5$\mu$m and much fainter at 5.8$\mu$m and even fainter at
8.0$\mu$m.

\subsection{Dust Temperature}
The hot dust in HII regions competes with the gas for stellar UV photons. It
also plays an important role in the processing of radiation and cooling of HII
regions.  It is therefore, of high interest to measure the temperature
distribution of the dust as a function of distance from the source(s) of
ionization. To do this, it is necessary to estimate what fraction of the
8$\mu$m emission interior to the inner face of the 8$\mu$m shell is due to
thermal continuum emission as opposed to PAH emission from the front and back
sides of the shell.  We assume that the PAH emission is confined to a
spherical shell of inner radius r and thickness dr, that the PAH emission is
proportional to the line-of-sight depth through the shell (front and back) and
that emission is homogeneous.  With these assumptions, the path length S
through the shell at each impact parameter y (which, for objects far from the
observer, is directly proportional to the angular separation) is:

\begin{eqnarray}
S &=& 2\left(\sqrt{r^2+2r dr +dr^2-y^2}-\sqrt{r^2-y^2}\right)
\end{eqnarray}

\begin{figure}
\caption{N49: Observed, azimuthally averaged, radial profile of 8 $\mu$m
  emission (+) compared with a model of shell emission (solid line) normalized
  to the observed brightness at the center of N49.}
\label{n49dust}
\epsscale{1}
\plotone{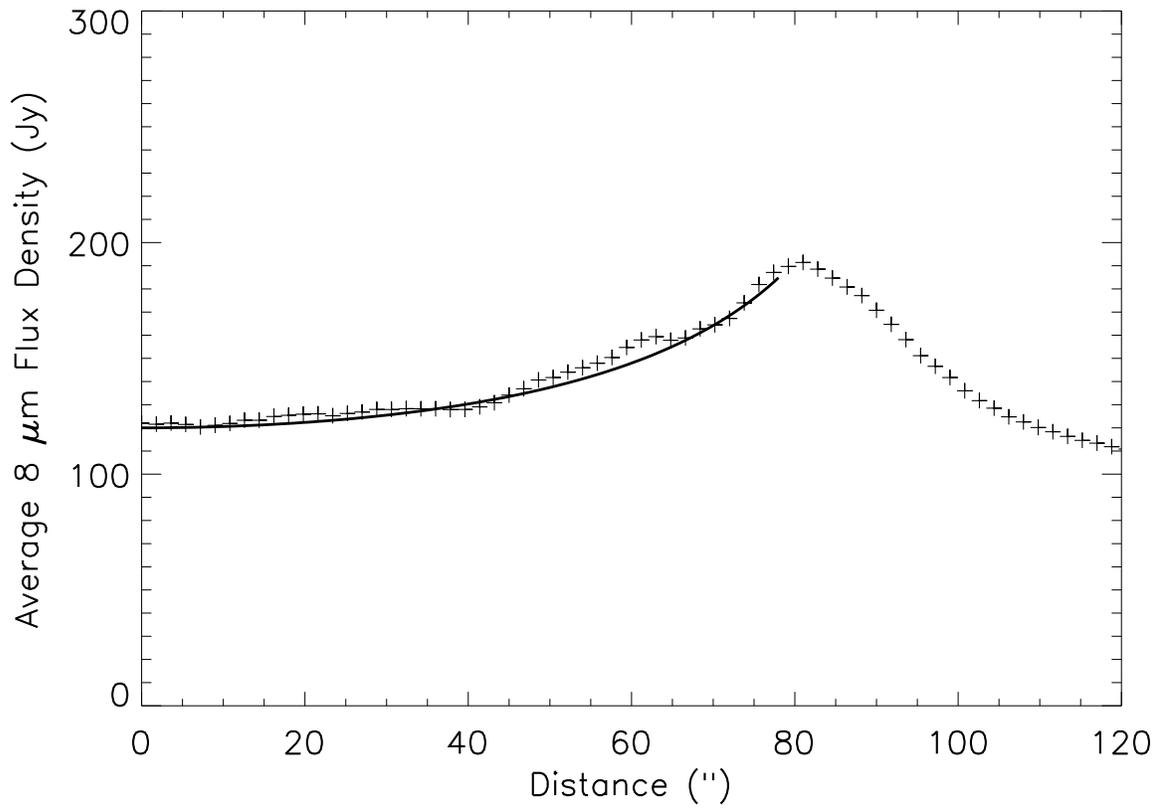}
\end{figure}

Figure \ref{n49dust} shows the azimuthally averaged observed radial 8$\mu$m
brightness of N49 with the profile predicted by eq.(1) normalized to the
observed flux density at the center of the bubble and superimposed on the
observations.  Surprisingly, the profile predicted by the assumed simple
geometry fits the observations very well.  A check on this assumption is the
fact that there is no apparent dip in brightness at 8$\mu$m at the center of
N49 but is very apparent at 24$\mu$m and 20 cm.  Taking the good fit of the
predicted profile with the observations at face value implies that almost all
the 8$\mu$m emission interior to the shell is due to PAH emission from the
front and back faces of the shell.  Only from about 45'' to 60'' is there
possibly a measurable excess that might be claimed to be thermal continuum
emission.  If this is the thermal component at 8$\mu$m, it is too small to
measure accurately enough to determine the dust temperature.  Using the noise
level of the 8$\mu$m emission at the center of the bubble along with the
measured 24$\mu$m emission, we can only place an upper-limit ($<$ 150 K) on
the interior dust temperature.

The turndown in 8$\mu$m brightness beginning at $\sim$ 80'' outward indicates
the decline in brightness of the PDR.  It declines at larger radii because of
dilution of stellar photons that can excite the PAH features.  We have not
done a similar analysis for N21 because of its lack of symmetry.

In N10, on the other hand, there appears to be 8 $\mu$m and 5.8 $\mu$m
emission coincident with the central 20 cm peak that is significantly stronger
than the shell emission (see Figure \ref{n10lat2}). This cannot be front or
back-face emission, as in N49, but rather it is likely thermal emission from
hot dust.  Unfortunately, the 24 $\mu$m emission is saturated.  Using only the
IRAC bands, we can only place an upper-limit ($<$250 K) on the interior dust
temperature.

\subsection{Morphology Comparison}
It is instructive to compare the MIR morphology of these three bubbles since
they are each unique in interesting ways. The most important difference is
that in N10 and N21, 24 $\mu$m and 20 cm emission is centrally peaked, whereas
in N49 there is a cavity. That is, the driving star in N49 appears to have
evacuated its immediate surroundings of hot dust (as traced by 24 $\mu$m
emission) and gas (as traced by 20 cm emission). This difference in hot dust
structure could be caused by either N49 containing an earlier ionizing star
(O5V vs. O6.5V, see \S 5) or by a difference in age (N49 being more evolved
than N10). A second difference is the detection of 8 $\mu$m emission toward
the center of N10 and N21 but its absence toward the center of N49. This
difference implies that the dust at the center of N10's and N21 is hotter than
in N49.

Since even cooler O stars destroy PAHs in their neighborhood (\S 4.1), the 8
$\mu$m emission at the centers of N10 and N21 is likely dominated by thermal
emission from hot dust.  Presumably, small dust grains that give rise to
stochastic emission are destroyed near O stars.  Assuming that the 24 and 8
$\mu$m emission at the center of N21 are due to thermal emission from large
dust grains, we find T$_{dust}$ $\sim$ 200 K.  A similar estimate cannot be
made for N10 because the 24 µm emission is saturated; however, using 5.8 and
3.5 $\mu$m emission, which probably samples hotter dust still closer to the
central star than in N21, we find T$_{dust}$ $\sim$380 K for N10.  The dust
temperatures in N21 and N10 are higher than in N49 even though N49 is ionized
by a hotter star because the dust is much closer to the central star(s) in N10
and N21 than in N49.

\section{Associated YSOs and Ionizing Stars}

In this section, we examine the stellar populations observed at NIR-MIR
wavelengths toward each bubble. We are interested in the YSOs associated with
each bubble, especially those appearing on the bubble rims that may have been
triggered by the expanding bubble, as well as the ionizing stars responsible
for producing each bubble.

To identify YSO candidates, we fit GLIMPSE Archive sources combined with 24
$\mu$m photometry from MIPSGAL images with spectral energy distributions
(SEDs) from a large, pre-computed grid of YSO models (Robitaille et al.\ 
2006).  The grid consists of 20,000 2-D Monte Carlo radiation transfer models
(Whitney et al.\ 2003a,b; Whitney et al.\ 2004) spanning a complete range of
stellar mass and evolutionary stage and output at 10 viewing angles (a total
of 200,000 SEDs).  The model fitting tool uses a fast $\chi^2$-minimization
fitting algorithm (Robitaille et al.\ 2007) and includes a grid of Kurucz
(1993) stellar photosphere SEDs. We can robustly distinguish between YSOs and
extincted photospheres of main-sequence and giant stars because YSOs require a
thermal emission component from circumstellar dust to reproduce the shapes of
their mid-IR excesses.  The concept of fitting SEDs from a large grid of
models was tested by fitting the SEDs of several known YSOs in Taurus and
deriving physical properties in agreement with previous determinations based
on other methods (Robitaille et al. 2007). This model-based approach does a
reliable job of identifying YSO candidates, and additionally provides
information on the physical properties of the YSOs.

In Table \ref{YSOs} we present the YSO candidates listed in order of ascending
Galactic longitude for each bubble. For each YSO candidate, the set of
well-fit models (numbering $N_{\rm fits}$) was selected on the basis of
$\chi^2$ according to
\begin{equation}\label{good}
  \chi^2-\chi^2_{\rm min} \le 2N_{\rm data},
\end{equation}
where $\chi_{\rm min}^2$ is the goodness-of-fit parameter of the best-fit
model and $4\le N_{\rm data}\le 8$ is the number of flux datapoints used for
the fit. From the set of well-fit models we construct the cumulative
probability distributions of the YSO parameters. These include mass $M_\star$
of the central star, total luminosity (stellar plus disk accretion) $L_{\rm
  TOT}$ of the YSO, and envelope accretion rate $\dot{M}_{\rm env}$. In Table
\ref{YSOs} we report the ``best'' values of these YSO parameters, defined as
the value of each parameter for which the slope of the cumulative probability
distribution is maximized. We include the minimum and maximum parameter values
representing 95\% confidence intervals.  Table \ref{YSOs} includes the most
probable evolutionary stage of the YSO, a classification of the models
introduced by Robitaille et al.\ (2006) that parallels the observational
``Class'' taxonomy (Lada 1987).  A Stage I YSO is heavily embedded in its
infalling envelope, while a Stage II YSO is a more evolved disk-dominated
object. We do not detect any candidate Stage III objects, young stars with
remnant dust disks. This may be a selection effect, however, because the SEDs
of such sources tend to most closely resemble stellar photospheres, and we
have been conservative in our selection of YSO candidates.  The comments
column of Table \ref{YSOs} gives the apparent location of the YSO: (Rim) YSO
on the projected rim of the bubble and hence a possible example of triggered
star formation; (IRDC) YSO within an infrared dark cloud; (Bub) YSO inside the
bubble in projection; and (PDR) YSO within the bright diffuse PAH emission
from the PDR of the bubble.  We also include a [4.5] flag for two YSO
candidates that appear to be associated with bright, extended emission at 4.5
$\mu$m.  Excess emission in IRAC [4.5] probably indicates shocked molecular
(H$_2$ or CO) outflow or jet, a signpost of the early stages of massive star
formation (Smith et al. 2006; Davis et al.2007; Shepherd et al. 2007).

\begin{figure}
  \plotone{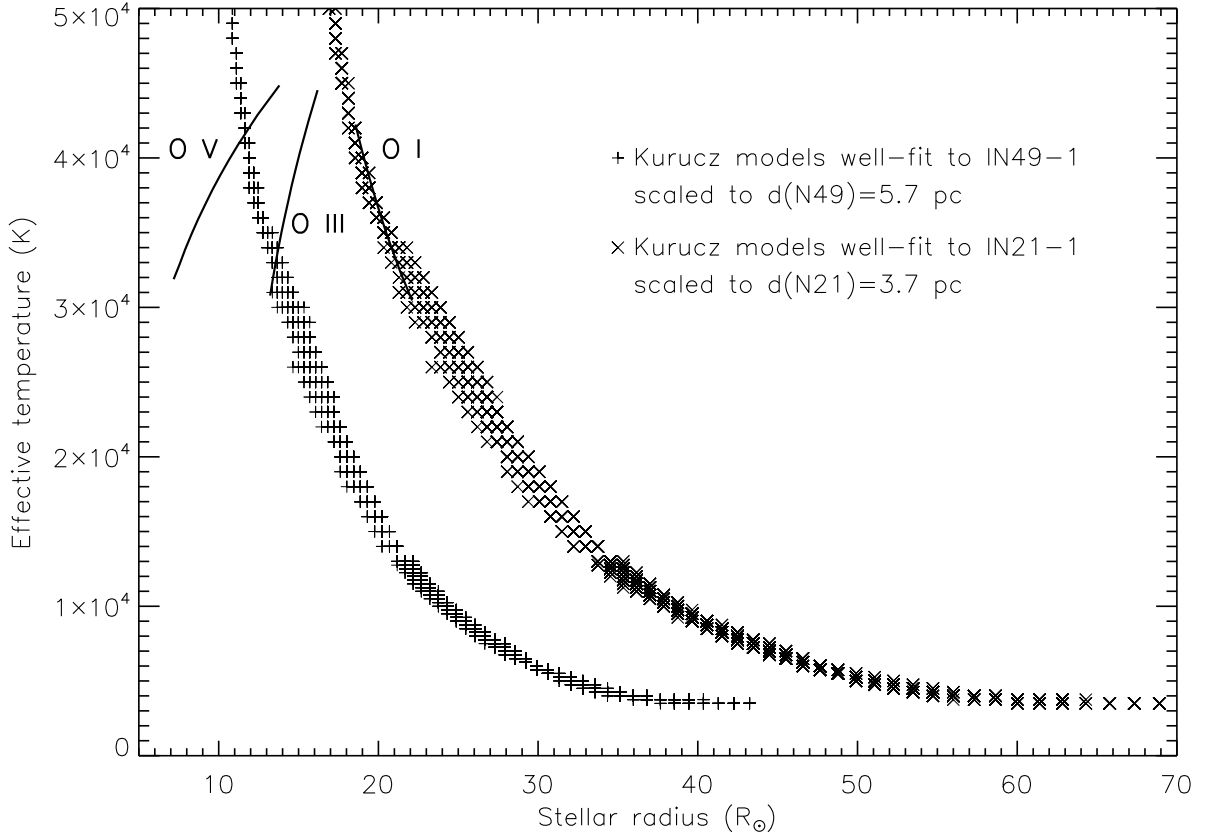}
  \caption{Examples of spectral type determinations for candidate ionizing  
    stars. The set of well-fit Kurucz (1993) stellar atmosphere models to the
    NIR--MIR SEDs of the best candidate ionizing stars in N49 and N21 have
    been plotted in $T_{\rm eff}$--$R$ space, where $R$ is calculated by
    scaling the model fits to the distance of their respective bubbles. The
    O-star $T_{\rm eff}$--$R$ relations of Martins et al. (2005) are
    overplotted as heavy curves for dwarfs, giants, and supergiants. The
    curves extend from types O3 to O9.5 for each luminosity class. From the
    intersections of the curves with the loci of model fits, a spectral type
    of O5 V or late O III is consistent with the observed fluxes of IN49-1.
    The spectral type of IN21-1 is degenerate among O supergiants. These
    degeneracies in spectral type can be lifted by considering the ionizing
    flux required to produce the \ion{H}{2} regions in which the stars are
    located. \label{fitter}}
\end{figure}

Identifying each bubble's ionizing stars is complicated by the fact that such
stars, while characteristically luminous, have reddened Rayleigh-Jeans
spectral slopes at NIR--MIR wavelengths. While the stellar photosphere models
included in the fitting tool are primarily intended to facilitate the
separation of YSO candidates from field stars, it is possible to use the
results of the photosphere fits to identify candidate ionizing stars lying
within the bubbles. Robitaille et al.\ (2007) have incorporated into the
fitting tool an interstellar extinction model using the MIR extinction
properties derived by Indebetouw et al. (2005).  To identify candidate hot
stars in the bubble, we forced the minimum extinction used by the fitting tool
to be 8--10 mag and selected sources for which the best-fit Kurucz model had a
photospheric temperature $T_{\rm eff}>20,000$ K. This revealed ${\sim}10$
candidate ionizing stars within the boundaries of each bubble. We then fit
these sources a second time, allowing the extinction to range from 0--30 mag.
The resulting set of well-fit models, selected according to Equation
\ref{good}, produces a tight curve of $T_{\rm eff}$ versus stellar radius $R$
when scaled to the distance of the bubble. This curve intersects with the
theoretical O-star $T_{\rm eff}$--$R$ relations of MSH05 (see Fig.
~\ref{fitter}).  This selects a particular group of models, providing
estimates of both the spectral types and the extinction toward each of the
candidate ionizing stars in each bubble.  The results of this analysis are
presented in Table \ref{ionizers}. The candidate stars are listed in
decreasing order of their likely importance in ionizing the \ion{H}{2} regions
of the bubbles. The fifth column indicates our best estimate of the ionizing
star based on location, spectral type and radio continuum emission. It is
important to note that this method of assigning spectral type is not
definitive. Any one of the candidate ionizing stars could be a less reddened,
foreground, cooler main sequence star or red giant.  Geometrical arguments,
however, strongly support the identification of the best candidates as real O
stars responsible for producing the bubbles.

\subsection{N10}
\begin{figure}
  \plotone{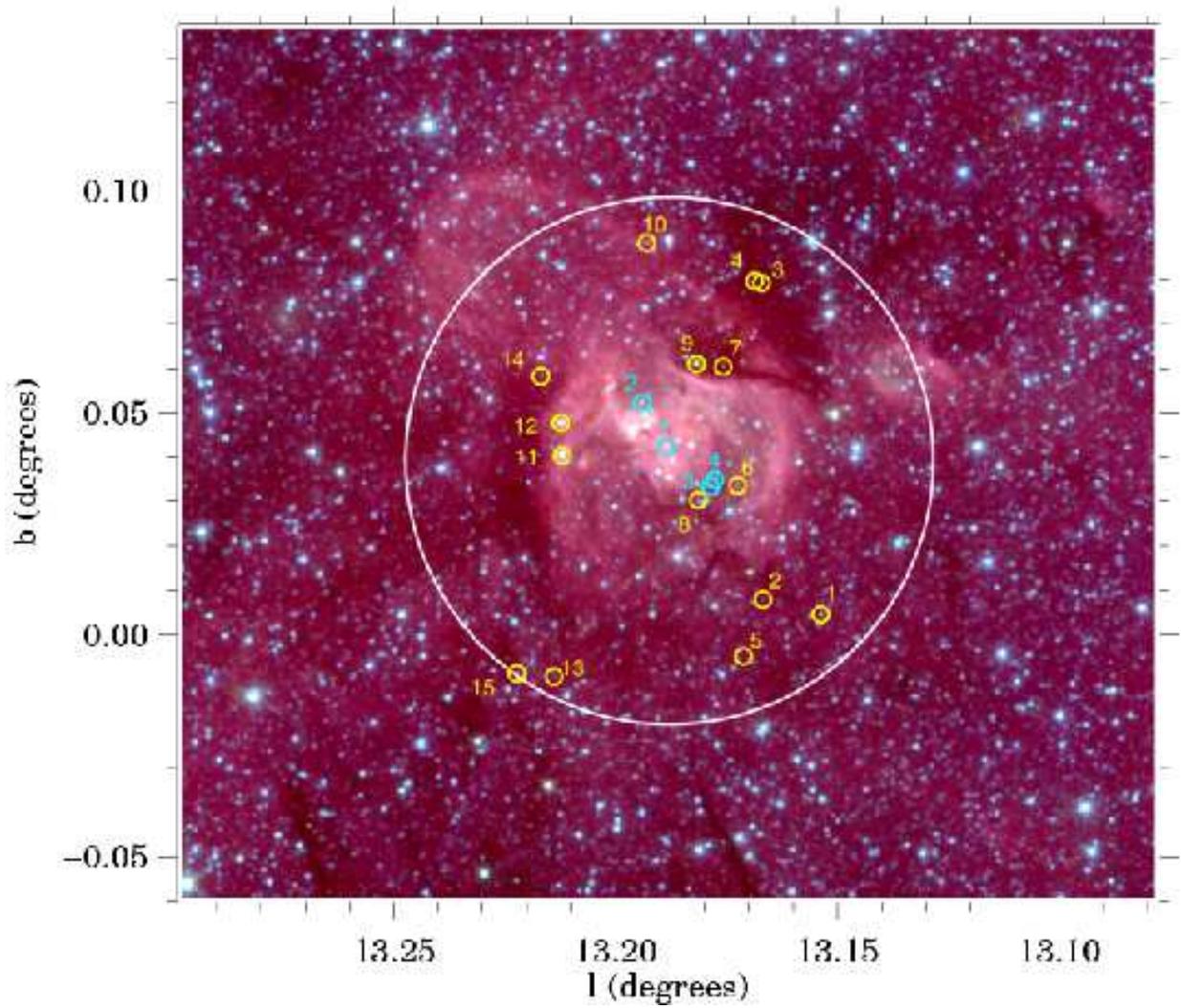}
  \caption{N10 image in 3.6 $\mu$m (blue), 4.5 $\mu$m (green) and 8.0 $\mu$m
    (red). Positions of YSO candidates in Table \ref{YSOs} and candidate
    ionizing stars in Table \ref{ionizers} are marked with yellow and cyan
    circles, respectively. The large white circle shows the area in which the
    GLIMPSE Point Source Archive was searched for YSO candidates.\label{N10}}
\end{figure}

Out of 687 GLIMPSE Archive sources analyzed within a 3.6' (5.2 pc) radius from
the center of N10, 15 were fit with high confidence by YSO models.  These YSO
candidates surround N10 (yellow circles in Figure \ref{N10}), and the
configuration is highly suggestive of triggered massive star formation.  The
bubble is bordered on 2 sides by infrared dark clouds (IRDCs).  4 candidate
highly-embedded Stage I massive YSOs appear to be located on the bubble rims
(sources N10-7, 5.8--16.7 M$_{\sun}$; N10-9, 8.3--13.5 M$_{\sun}$; N10-11,
8.4--12.7 M$_{\sun}$; and N10-12, 8.9--17.1 M$_{\sun}$, see Table \ref{YSOs}).
Two of these YSOs are also in an IRDC and the other two YSOs appear close to
an IRDC.

Four stars have been identified as possible ionizing stars located in
projection inside the bubble.  It should be emphasized that these stars are
well-fit with stellar photospheres and are not YSOs. The spectral energy
distributions of these stars suggest spectral types ranging from O7.5V to O6V
(see Table 2 and cyan circles in Figure 19).  Although the spectral types are
rather uncertain, if we take them at face value and use the models of MSH05,
their combined UV photon flux is $\sim$2.2x10$^{49}$ s$^{-1}$, more than
adequate to maintain ionization of the HII region (1.6$\times$10$^{49}$
s$^{-1}$, based on 20 cm emission). Some UV, however, is required to heat the
dust at the center of N10. IN10-1 appears to lie at the center of the bubble
and near the center of the radio and saturated diffuse 24 $\mu$m emission
filling the bubble, while IN10-2 appears to lie in a sub-cavity near a second
peak of bright diffuse MIR emission that may be hiding a cluster of later-type
OB stars or YSOs. Preliminary optical spectroscopy obtained using the Wyoming
Infrared Observatory (WIRO) indicates that IN10-3 is an O8-B0 V star
(Kobulnicky, priv. comm.). Thus, the above method appears to be a robust
method of identifying ionizing star candidates. The discrepancy between the
spectral identification and the analysis presented above may be caused by the
uncertain kinamtic distance used for the bubble.  The presence of the IRDCs on
2 sides of N10 suggest that this bubble may be density-bounded in those
directions, allowing the bubble to expand asymmetrically, with the center of
influence from these 2 stars offset from the geometric center of the bubble.


\subsection{N21}
\begin{figure}
  \plotone{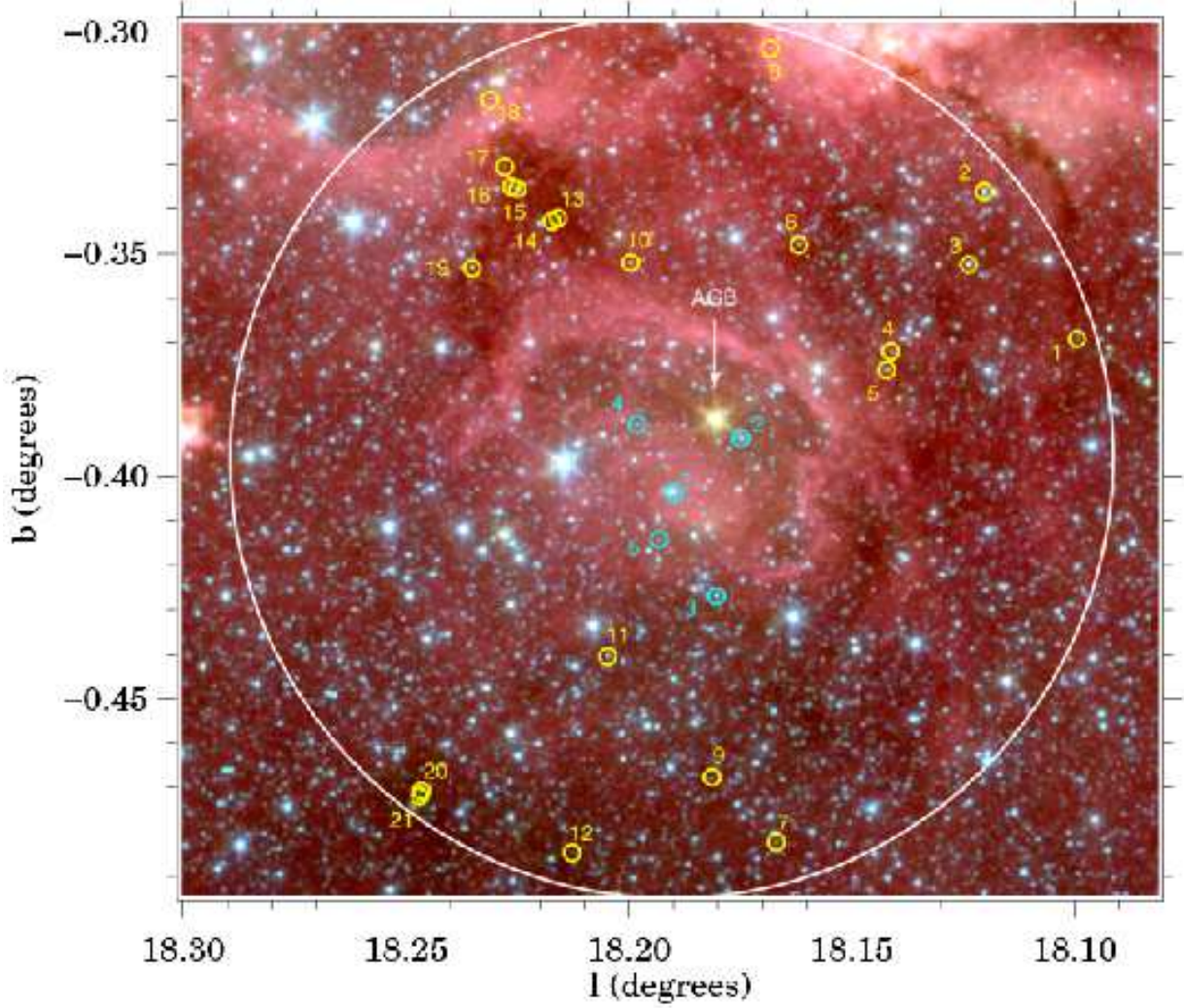}
  \caption{Same as Fig.\ \ref{N10}, but showing bubble N21. The
    MIR-bright asymptotic giant branch star is labeled. Although this star
    appears to lie inside the bubble, it is most likely a background AGB
    star.\label{N21}}
\end{figure}

N21 is associated with a larger HII region complex. While 21 YSO candidates
(yellow circles in Figure \ref{N21}) were selected from the 2333 GLIMPSE
Archive sources within 6' (6.5 pc, white circle) from the center of the
bubble, more YSO candidates probably lie beyond our search radius. The
greatest concentration of YSOs appears in the IRDC located midway between N21
and a neighboring bubble in the upper-left corner of Figure \ref{N21} (N22 in
the Churchwell et al.  2006 catalog), but it is unclear whether this IRDC is
physically associated with either of the bubbles.  Perhaps the most
interesting single point source in this image is the very bright greenish-red
source that appears to lie inside the bubble.  This source has an extremely
positive spectral index in the NIR--MIR (it is undetected in 2MASS $J$ and $H$
and saturated in all 4 GLIMPSE bands) and is bright enough at longer MIR
wavelengths to be detected in all 4 MSX bands and to saturate the MIPS 24
$\mu$m band. Although this spectral index is suggestive of a massive Class I
YSO, this source is undetected by MIPS at 70 $\mu$m, while a YSO SED should
peak near 70 $\mu$m. We can estimate ranges of extinction and distance
consistent with the source being an AGB star by assuming a typical K-magnitude
of -7.56 (Sohn et al. 2006). Assuming A$_V <$ 40 mag (larger than any source
detected in the field), the distance is $>$ 10 kpc, indicating a background
AGB star.


The best candidate star for producing N21 is IN21-1, which lies at the center
of the brightest radio and diffuse 24 $\mu$m emission. The model fits to this
star, when scaled to the 3.7 kpc kinematic distance of the \ion{H}{2} region,
produce a stellar radius too large to be a main-sequence star (see
Fig~\ref{fitter}). Instead, the models lie exactly along the $T_{\rm
  eff}$--$R$ curve derived by MSH05 for O supergiants, a highly suggestive
correspondence. Because the 2 curves overlap the spectral type is degenerate.
Preliminary optical spectroscopy obtained using WIRO indicates IN21-1 is a
late-O supergiant (Kobulnicky, priv. comm.), again confirming the method of
identifying ionizing star candidates. The ionizing photon flux required to
maintain the \ion{H}{2} region suggests an early BI.

\subsection{N49}
\begin{figure}
  \plotone{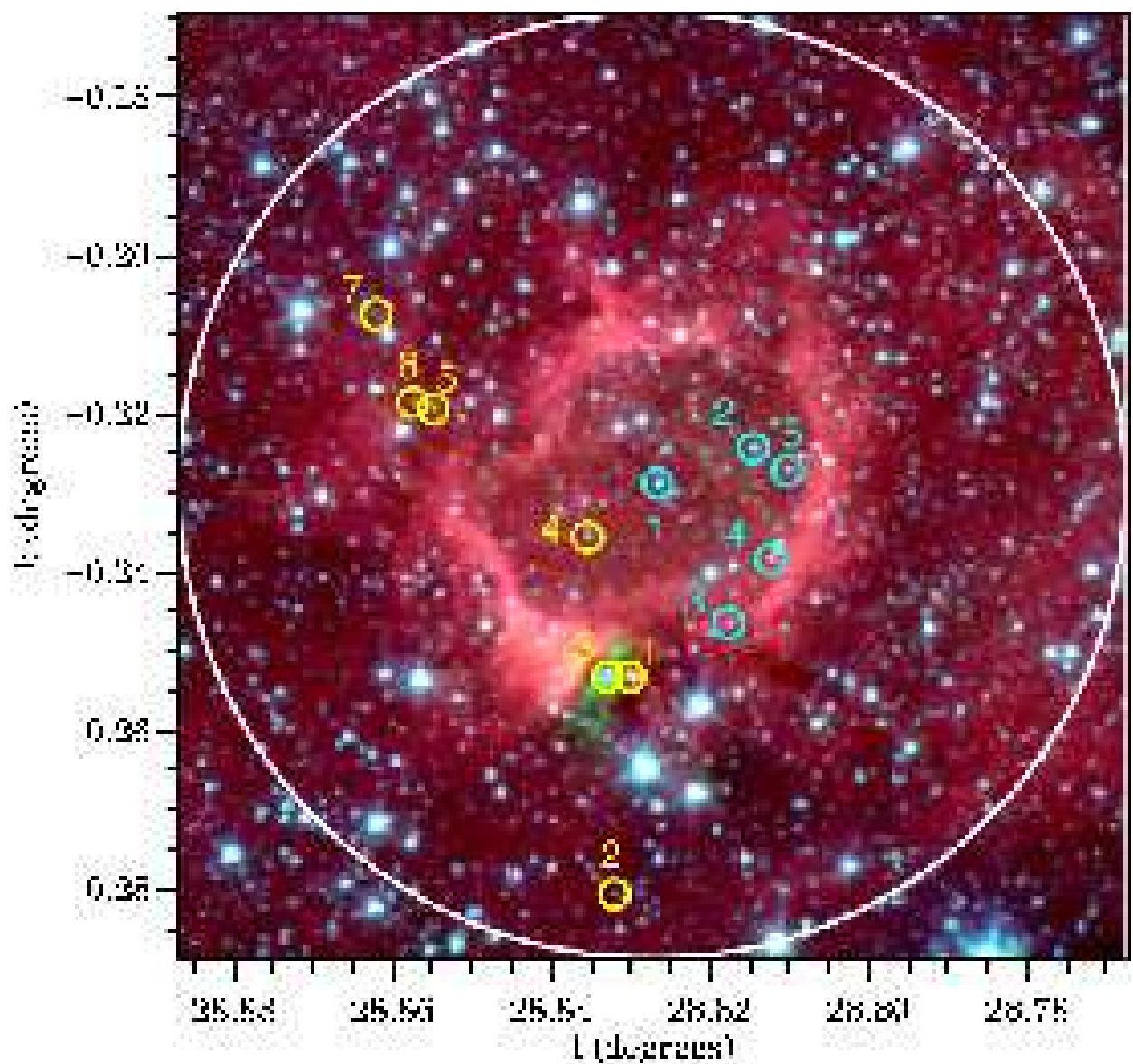}
  \caption{Same as Fig.\ \ref{N10}, but showing bubble N49.\label{N49}}
\end{figure}

\begin{figure}
  \plottwo{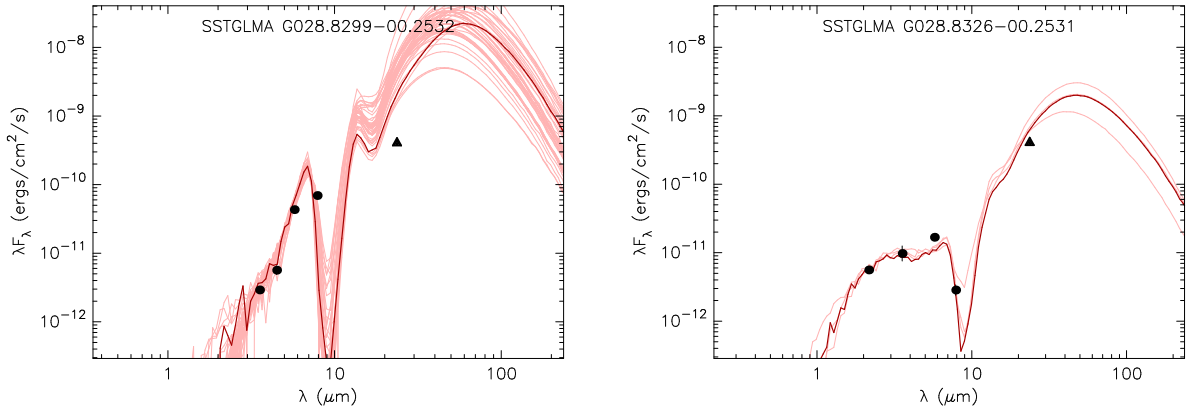}{f17b.eps}
  \caption{Model fits to the SEDs of 2 candidate massive YSOs
    appearing on the rim of bubble N49. Measured broadband fluxes and lower
    limits are plotted as heavy dots with error bars and triangles,
    respectively. The set of well-fit model SEDs are overplotted as curves,
    with the heavy curve showing the best fit. {\it Left panel:} YSO N49-1 is
    extremely red in the GLIMPSE bands. {\it Right panel:} YSO N49-3 is
    associated with the brightest extended 4.5 $\mu$m emission in our sample,
    indicative of a powerful molecular outflow. Because the 4.5 $\mu$m emission is
    extended, and the YSO models do not incorporate molecular line emission,
    this band was not used in fitting the SED of this object.  These sources
    are very close together, and their 24 $\mu$m emission is borderline
    confused and saturated, so lower limits were employed for the
    fitting.\label{N49YSOs}}
\end{figure}

Within a radius of 3.6' (6 pc) from the center of N49, 722 GLIMPSE Archive
sources were analyzed and 7 were fit with high confidence as YSOs (yellow
circles in Figure \ref{N49}). N49-1 (see Table \ref{YSOs} and Figure
\ref{N49YSOs}) was fit with Stage I models ranging in mass from 14 to 29
M$_{\sun}$, making it potentially the most massive YSO in our sample.  N49-3
sports a spectacular example of a bipolar outflow seen in 4.5 $\mu$m emission.
Both YSO candidates appear to lie in an IRDC just beyond the rim of the bubble
and adjacent to a bright knot of diffuse 8.0 $\mu$m and 24 $\mu$m emission that
probably hides additional YSOs undetected by GLIMPSE because they are masked
by the diffuse MIR emission.  Other potential YSOs in the IRDC may not have
been detected by GLIMPSE due to confusion and high extinction.  Hence N49-1
and N49-3 may be massive members of a YSO cluster that has been triggered by
the expansion of the bubble.

N49 contains a star, IN49-1, at the center of the wind-evacuated cavity
(Figure \ref{N49}). Given the strong circular symmetry observed in all bands,
this star is very likely the ionizing star responsible for producing the
bubble. IN49-1 is fit both by an O8 III and an O5 V star (see Fig
\ref{fitter}). The UV photon flux implied by the radio emission (10$^{48.89}$
s$^{-1}$) is in closer agreement with an O8 III star (10$^{48.88}$ s$^{-1}$)
than an O5 V star (10$^{49.22}$ s$^{-1}$). However, the age implied by the
presence of triggered star-formation is closer to that of an O5 V star
($\sim$10$^5$ yrs) than that of an O8 III star (10$^{6}$ yrs), so we favor the
O5 V classification. Presumably the excess UV is responsible for heating the
dust that produces the 24 $\mu$m emission inside the HII region. The model
SEDs fit to the 2 brightest sources (N49-1 and N49-3) were based upon the 4
IRAC bands plus a 24 $\mu$m lower limit. The 24 $\mu$m lower limit effectively
places lower limits on the YSO masses and luminosities, but the upper limits
are difficult to constrain based upon the available photometry.

The central cavity in N49 seems to indicate that this bubble is stellar
wind-dominated and therefore should be describable by the analytic relations
of Weaver et al. (1977):

\begin{eqnarray*}
R(t) &\propto &n_o^\frac{-1}{5} L_w^\frac{1}{5} t^\frac{3}{5}
\end{eqnarray*}
where R(t) is the radius, n$_o$ is the initial ambient density, L$_w$ is the
stellar wind luminosity, and t is the dynamical age of the bubble.  From this,
one can estimate the age of a bubble as a function of no given the wind
luminosity and a measured radius.  We use the stellar parameters of MSH05 to
estimate the mass loss rate (1.5 x 10$^{-6}$ M$_\odot$ yr$^{-1}$) and wind
luminosity of the ionizing star ($\sim$4 x 10$^{36}$ erg s$^{-1}$) using the
prescription of Vink, de Koter, and Lamers (2001).  Having the wind luminosity
and measured bubble radius, we can estimate the age of the bubble for an
assumed ambient density.  Several Class I or younger YSOs have been identified
along the periphery of the bubble.  Assuming that the YSOs were triggered by
the expansion of the bubble would require the bubble to be at least 10$^{5}$
yr old because this is the time scale for massive YSO formation.
Using the estimated initial ambient density of $\ge$10$^3$ cm$^{-3}$, the
Weaver et al.  (1977) relations implies a dynamical age $\ge$10$^{5}$ yr, a
very young bubble but easily old enough to have spawned a second generation of
star formation.  Freyer et al. (2006), however, have shown using numerical
simulations that the energy transfer efficiency may be lower than assumed in
analytic models.  If this is the case, then the age estimates based on the
analytic relations will be under-estimated, giving even more time for
triggered star formation to occur.

\section{Conclusions}

Based on MIR and 20 cm observations of three bubbles, we conclude the
following:\\
$\bullet$ At the center of each bubble there is a region containing ionized
gas and hot dust delineated by radio free-free emission and 24 $\mu$m
emission, respectively.\\
$\bullet$ At the center of N10 and N21 (but not N49), the hot dust is also
traced by 5.8 and 8.0 $\mu$m emission.\\
$\bullet$ Based on the decreased ratio of 8.0 $\mu$m/4.5 $\mu$m, 5.8
$\mu$m/4.5 $\mu$m and the lack of a similar decrease in 8.0 $\mu$m/3.6 $\mu$m,
we conclude that inside the 8 $\mu$m shell PAHs are destroyed by hard, direct
stellar UV radiation.\\
$\bullet$ Based on the increased 8.0 $\mu$m/4.5 $\mu$m, 5.8 $\mu$m/4.5 $\mu$m,
the bright 8.0 $\mu$m ring emission is dominated by PAH emission that defines
the PDR region around each bubble.\\
$\bullet$ We have identified YSO candidates and probable ionizing sources for
each bubble. This was accomplished by employing the SED model fitter developed
by Robitaille et al. (2007) to fit model SEDs of YSOs and hot stellar
photospheres (Kurucz 1993) to the observed fluxes.\\
$\bullet$ Based on morphology and environment, several of the identified YSO
candidates in N10 and N49 appear to be triggered by expansion of the
bubbles.\\
$\bullet$ The wind-blown cavity at the center of N49 appears to be produced by
a central 05 V star, the hottest ionizing star observed here. This bubble
appears to be dominated by the wind from the O5 star and have a dynamical age
of $\ge$10$^5$ yrs.\\

\acknowledgements 
We would like to acknowledge Chip Kolbunicky for obtaining
optical spectra of the candidate ionizing stars. An anonymous referee made
many comments which improved the paper. E.B.C would like to acknowledge
support through NASA contract \# 1275394.

\clearpage


%

\clearpage

%

%
%

\clearpage

\clearpage

\clearpage

\begin{deluxetable}{lcrrrrcccccccccrr}
\tabletypesize{\tiny}
\rotate
\tablecaption{Model Parameters for YSO Candidates Associated with the
  Bubbles \label{YSOs}}

\tablewidth{0pt}
\tablehead{
  \multicolumn{3}{c}{ } & \multicolumn{3}{c}{$M_{\star}$ (M$_{\sun}$)} & \multicolumn{3}{c}{$L_{\rm TOT}$ (L$_{\sun})$} & \multicolumn{3}{c}{$\dot{M}_{\rm env}$ (M$_{\sun}$ yr$^{-1}$)} & \multicolumn{2}{c}{ } \\ 
  \colhead{ID} & \colhead{Name (G$l+b$)} & \colhead{$N_{\rm fits}$} & \colhead{best} & \colhead{min} & \colhead{max} & \colhead{best} & \colhead{min} & \colhead{max}  & \colhead{best} & \colhead{min} & \colhead{max} & Stage & Comments\tablenotemark{a} \\
}
\startdata
N10-1 &  G013.1536+00.0040 &   86 &    .7 &    2.9 &    7.3 &    315 &     76 &    532 &   2.3E-04 &  4.1E-05 &  9.4E-04 & I &  \\
N10-2 &  G013.1667+00.0074 &   12 &    .1 &    1.0 &    5.5 &     26 &     15 &    682 &   1.5E-04 &  0.0E+00 &  4.1E-04 & I &  \\
N10-3 &  G013.1670+00.0794 &    7 &    .6 &    0.9 &    3.0 &     43 &     33 &    128 &   3.3E-04 &  2.1E-04 &  4.1E-04 & I & IRDC \\
N10-4 &  G013.1686+00.0798 &    3 &   1.7 &    3.7 &    3.7 &    107 &    107 &    107 &   2.1E-04 &  2.1E-04 &  2.1E-04 & I & IRDC \\
N10-5 &  G013.1711-00.0055 &  351 &   3.4 &    0.3 &    7.1 &     35 &      9 &    766 &   0.0E+00 &  0.0E+00 &  3.8E-04 & II &  \\
N10-6 &  G013.1725+00.0333 &  597 &    .1 &    0.7 &    7.5 &    152 &     12 &    600 &   2.2E-05 &  1.6E-06 &  6.0E-04 & I &  Bub \\
N10-7 &  G013.1758+00.0604 &   57 &    .4 &    5.8 &   16.7 &   6479 &    273 &   8570 &   8.9E-04 &  2.2E-04 &  1.6E-03 & I & Rim, IRDC \\
N10-8 &  G013.1814+00.0300 & 2395 &   3.6 &    0.9 &   11.1 &    246 &     24 &   3225 &   0.0E+00 &  0.0E+00 &  1.4E-03 & I & Bub \\
N10-9 &  G013.1818+00.0610 &  197 &    .3 &    0.4 &   13.5 &    2820 &    142 &  13880 &   1.0E-03 &  3.7E-06 &  2.1E-03 & I &  Rim, IRDC, [4.5] \\
N10-10 &  G013.1931+00.0885 &   84 &   5.1 &    1.9 &    6.2 &    440 &     62 &    440 &   2.2E-03 &  2.2E-05 &  2.2E-03 & I & PDR \\
N10-11 &  G013.2124+00.0401 &   42 &   0.7 &    8.4 &   12.7 &   3858 &   2170 &   6900 &   9.3E-04 &  1.5E-04 &  1.0E-03 & I & Rim \\
N10-12 &  G013.2127+00.0476 &  244 &   8.9 &    8.9 &   17.1 &   5505 &   3342 &  29590 &   2.2E-04 &  6.4E-05 &  5.8E-03 & I & Rim \\
N10-13 &  G013.2142-00.0101 &   27 &   3.6 &    2.5 &    4.2 &    206 &     44 &    258 &   0.0E+00 &  0.0E+00 &  0.0E+00 & II &  \\
N10-14 &  G013.2172+00.0582 & 2400 &   4.2 &    1.4 &    8.7 &    144 &     33 &   1349 &   0.0E+00 &  0.0E+00 &  4.6E-04 & I & IRDC \\
N10-15 &  G013.2227-00.0093 &   13 &   0.5 &    0.2 &    0.5 &     16 &     15 &     20 &   2.9E-06 &  2.3E-06 &  3.9E-06 & I & IRDC \\
\tableline
N21-1 &  G018.0981-00.3692 & 1302 &   .8 &    0.4 &    4.6 &     27 &      6 &     87 &   1.3E-05 &  1.2E-06 &  2.3E-04 & I & IRDC \\
N21-2 &  G018.1192-00.3359 & 2212 &   .3 &    3.2 &    6.9 &    314 &     78 &   1563 &   0.0E+00 &  0.0E+00 &  4.3E-05 & II & PDR \\
N21-3 &  G018.1227-00.3524 &   49 &   .4 &    4.6 &    6.8 &   1151 &    383 &   1543 &   0.0E+00 &  0.0E+00 &  0.0E+00 & II & PDR \\
N21-4 &  G018.1403-00.3721 & 1581 &   .5 &    1.5 &    8.6 &    102 &     30 &   1216 &   0.0E+00 &  0.0E+00 &  5.4E-04 & I & PDR \\
N21-5 &  G018.1412-00.3763 & 1166 &   .9 &    2.6 &    5.8 &    163 &     76 &    719 &   0.0E+00 &  0.0E+00 &  1.4E-04 & II & PDR \\
N21-6 &  G018.1611-00.3480 &  354 &   .1 &    2.6 &    4.2 &     80 &     42 &    141 &   0.0E+00 &  0.0E+00 &  1.3E-07 & II & PDR \\
N21-7 &  G018.1662-00.4833 &  799 &   .6 &    1.6 &    4.0 &     44 &     16 &    197 &   0.0E+00 &  0.0E+00 &  1.1E-08 & II &  \\
N21-8 &  G018.1677-00.3034 &   60 &   .5 &    1.2 &    8.0 &    474 &     35 &    903 &   5.0E-05 &  1.0E-05 &  2.9E-04 & I &  PDR \\
N21-9 &  G018.1808-00.4686 & 3546 &   .8 &    0.5 &    4.7 &     58 &      8 &    140 &   0.0E+00 &  0.0E+00 &  2.3E-04 & I &  \\
N21-10 &  G018.1992-00.3520 &  276 &  4.2 &    3.2 &    5.2 &    258 &    144 &    583 &   0.0E+00 &  0.0E+00 &  0.0E+00 & II &  PDR \\
N21-11 &  G018.2044-00.4412 & 1015 &  2.9 &    0.9 &    5.1 &     59 &     26 &    143 &   0.0E+00 &  0.0E+00 &  3.4E-04 & I &  \\
N21-12 &  G018.2125-00.4857 &   25 &  1.0 &    0.5 &    2.0 &     11 &      7 &     37 &   2.4E-05 &  6.8E-06 &  4.7E-04 & I &  \\
N21-13 &  G018.2157-00.3419 &   33 &  8.6 &    1.2 &    8.6 &   1443 &     62 &   1735 &   1.3E-03 &  1.2E-05 &  1.3E-03 & I &  IRDC \\
N21-14 &  G018.2171-00.3426 &   83 &  0.5 &    0.2 &    6.4 &     16 &     16 &    379 &   2.9E-06 &  2.3E-06 &  1.3E-03 & I &  IRDC \\
N21-15 &  G018.2249-00.3352 &  371 &  6.1 &    1.1 &    6.1 &    358 &     19 &    358 &   6.4E-05 &  0.0E+00 &  5.8E-04 & I &  IRDC \\
N21-16 &  G018.2262-00.3348 &   63 &  1.3 &    0.4 &    2.4 &     44 &     34 &    134 &   6.7E-06 &  4.3E-06 &  3.7E-05 & I &  IRDC \\
N21-17 &  G018.2277-00.3303 & 3136 &  1.5 &    0.7 &    5.2 &     69 &     12 &    154 &   0.0E+00 &  0.0E+00 &  3.0E-04 & I &  IRDC \\
N21-18 &  G018.2311-00.3150 & 5360 &  1.7 &    0.7 &    5.7 &     18 &     10 &    546 &   0.0E+00 &  0.0E+00 &  2.9E-04 & I &  PDR \\
N21-19 &  G018.2351-00.3532 &   83 &  5.1 &    0.4 &    6.0 &    182 &     26 &    480 &   9.4E-04 &  0.0E+00 &  9.4E-04 & I &  IRDC \\
N21-20 &  G018.2466-00.4717 &   20 &  2.9 &    1.2 &    4.1 &    139 &     56 &    235 &   1.3E-03 &  0.0E+00 &  1.3E-03 & I &  \\
N21-21 &  G018.2470-00.4728 &   19 &  5.5 &    2.0 &    6.0 &    161 &     46 &    194 &   6.3E-04 &  1.5E-04 &  9.9E-04 & I &  \\
\tableline
N49-1 &  G028.8299-00.2532 &   47 &    9.0 &   14.0 &   29.0 &  30040 &   5251 & 102400 &   8.9E-04 &  4.1E-04 &  6.7E-03 & I &  Rim, IRDC \\
N49-2 &  G028.8318-00.2808 & 1461 &   34.0 &    1.6 &    4.7 &     60 &     19 &    205 &   0.0E+00 &  0.0E+00 &  1.7E-05 & II &  \\
N49-3 &  G028.8326-00.2531 &    4 &    0.5 &    9.3 &   12.7 &   2942 &   2288 &   3657 &   3.4E-04 &  1.2E-04 &  3.4E-04 & I &  Rim, IRDC,[4.5] \\
N49-4 &  G028.8352-00.2354 & 1578 &   33.7 &    2.8 &    7.9 &    207 &    101 &   1050 &   0.0E+00 &  0.0E+00 &  7.0E-04 & I &  Bub \\
N49-5 &  G028.8547-00.2192 &   72 &    1.3 &    0.3 &    8.1 &     44 &     34 &    924 &   6.7E-06 &  4.0E-06 &  1.0E-03 & I &  PDR \\
N49-6 &  G028.8573-00.2184 &  276 &    0.5 &    0.2 &    6.9 &     16 &     15 &    379 &   2.9E-06 &  2.3E-06 &  1.2E-03 & I &  PDR \\
N49-7 &  G028.8619-00.2072 &  812 &    2.3 &    0.7 &    3.9 &     18 &      7 &     35 &   1.2E-06 &  8.6E-08 &  9.3E-05 & I &  \\
\enddata
\tablenotetext{a}{The flags in the comments column are as follows:
  Rim=source on the rim of the bubble; IRDC=source within an infrared
  dark cloud; PDR=source within bright diffuse PAH background emission
  in the photodissociation region; and [4.5]=source exhibits extended
  excess emission at 4.5 $\mu$m.}
\end{deluxetable}

\clearpage
\begin{deluxetable}{lcccc}
  \tablecaption{Candidate Ionizing Stars\label{ionizers}}
  \tablewidth{0pt}
  \tablehead{\colhead{ID} & \colhead{Name (G$l+b$)} &
    \colhead{Spectral Type} & \colhead{$A_V$} & \colhead{Best?} \\
    }
    \startdata
    IN10-1 & 13.1887+00.0421 & O7.5 V & 7 & \checkmark \\
    IN10-2 & 13.1942+00.0521 & O6.5 V & 7 &  \\
    IN10-3 & 13.1786+00.0331 & O6 V & 5 & \checkmark \\
    IN10-4 & 13.1777+00.0346 & O7 V & 8 & \\
    \tableline
    IN21-1 & 18.1893-00.4041 & early B I\tablenotemark{a} & 6 & \checkmark \\
    IN21-2 & 18.1742-00.3918 & O6 V & 9 & \\
    IN21-3 & 18.1798-00.4275 & O8 V & 7.5 & \\
    IN21-4 & 18.1977-00.3886 & O8.5 V & 13 & \\
    IN21-5 & 19.1928-00.4147 & early B V& 8 & \\
    \tableline
    IN49-1 & 28.8263-00.2287 & O5 V & 10.5 & \checkmark \\
    IN49-2 & 28.8142-00.2241 & O5.5 V & 7.5 & \\
    IN49-3 & 28.8174-00.2464 & O7 V & 7.5 & \\
    IN49-4 & 28.8119-00.2383 & O9 & 10 & \\
    IN49-5 & 28.8098-00.2270 & B0 & 6 & \\
    \enddata
    \tablenotetext{a}{For this star, the observed $T_{\rm eff}$--$R$ relation
      at the 3.7-kpc kinematic distance of N21 overlaps with the MSH05 curve
      for O supergiants, so the spectral type is degenerate. We have assigned
      a spectral type based upon the Lyman continuum photon flux required to
      ionize the \ion{H}{2} region in N21.}
\end{deluxetable}


\begin{thebibliography}{}
\bibitem[Benjamin et al.(2003)]{2003PASP..115..953B} Benjamin, R.~A., et 
al.\ 2003, \pasp, 115, 953 

\bibitem[Carey et al.(2005)]{2005AAS...207.6333C} Carey, S.~J., et al.\ 
2005, Bulletin of the American Astronomical Society, 37, 1252 

\bibitem[aa]{aa}Castor, J., McCray, R., \& Weaver, R. 1975, ApJ, 200, L107

\bibitem[baa]{baa} Capriotti, E.R. \& Kozminski, J.F. 2001, PASP, 113, 677

\bibitem[Churchwell \& GLIMPSE Team(2001)]{2001AAS...198.2504C} Churchwell, 
E., \& GLIMPSE Team 2001, Bulletin of the American Astronomical Society, 
33, 821 

\bibitem[Churchwell et al.(2006)]{2006ApJ...649..759C} Churchwell, E., et 
al.\ 2006, \apj, 649, 759 

\bibitem[Churchwell et al.(2007)]{2006ApJ...649..759C} Churchwell, E., et 
al.\ 2007, \apj, submitted

\bibitem[Cohen et al.(2007)]{2007MNRAS.374..979C} Cohen, M., et al.\ 2007, 
\mnras, 374, 979


\bibitem[Davis et al.(2007)]{CD07} Davis, C. J., Kumar, M. S. N.,
  Sandell, G., Froebrich, D., Smith, M. D., \& Currie, M. J. 2007,
  \mnras, 374, 29  

\bibitem[Deharveng et al.(2005)]{2005A&A...433..565D} Deharveng, L., 
Zavagno, A., \& Caplan, J.\ 2005, \aap, 433, 565 

\bibitem[Draine(2003)]{2003ARA&A..41..241D} Draine, B.~T.\ 2003, \araa, 41, 
241 

\bibitem[asdf]{asdf}Fazio, G.G., et al. 2004 ApJS, 154, 87

\bibitem[Freyer et al.(2003)]{2003ApJ...594..888F} Freyer, T., Hensler, G., 
\& Yorke, H.~W.\ 2003, \apj, 594, 888 

\bibitem[aa]{aa}Freyer, T., Hensler, G., Yorke, H.W. 2006, ApJ, 628, 262 [FHY06]

\bibitem[Garcia-Segura \& Franco(1996)]{1996ApJ...469..171G} Garcia-Segura, 
G., \& Franco, J.\ 1996, \apj, 469, 171 

\bibitem[Hoare(1990)]{1990MNRAS.244..193H} Hoare, M.~G.\ 1990, \mnras, 244, 
193 

\bibitem[Hoare et al.(1991)]{1991MNRAS.251..584H} Hoare, M.~G., Roche, 
P.~F., \& Glencross, W.~M.\ 1991, \mnras, 251, 584 
\bibitem[Hoare et al.(2007)]{2007prpl.conf..181H} Hoare, M.~G., Kurtz, 
S.~E., Lizano, S., Keto, E., \& Hofner, P.\ 2007, Protostars and Planets V, 
181 

\bibitem[Helfand et al.(2006)]{2006AJ....131.2525H} Helfand, D.~J., Becker, 
R.~H., White, R.~L., Fallon, A., \& Tuttle, S.\ 2006, \aj, 131, 2525 
\bibitem[Indebetouw et al.(2005)]{IM05} Indebetouw, R. et al.\ 2005,
  \apj, 619, 931
\bibitem[Kassis et al.(2006)]{2006ApJ...637..823K} Kassis, M., Adams, 
J.~D., Campbell, M.~F., Deutsch, L.~K., Hora, J.~L., Jackson, J.~M., \& 
Tollestrup, E.~V.\ 2006, \apj, 637, 823 

\bibitem[Kessler et al.(1996)]{1996A&A...315L..27K} Kessler, M.~F., et al.\ 
1996, \aap, 315, L27 

\bibitem[aa]{aa}Kr\"oger, D., Hensler, G., Freyer, T., 2006 A\&A, 450, L5

\bibitem[aa]{aa}Kr\"oger, D., Freyer, T., Hensler, G., Yorke, H.W. 2007 A\&A, submitted

\bibitem[Kurucz(1993)]{1993yCat.6039....0K} Kurucz, R.~L.\ 1993, VizieR 
Online Data Catalog, 6039, 0 
\bibitem[Lada(1987)]{CL87} Lada, C. J.  1987, in IAU Symp. 115: Star
  Forming Regions, eds. M. Peimbert \& J. Jugaku, 1

\bibitem[aa]{aa}Martins, F., Schaerer, D., Hillier, D.J. 2005, A\&A, 436, 1049

\bibitem[Peeters et al.(2003)]{2003asdu.confE..42P} Peeters, E., Tielens, 
A.~G.~G.~M., Allamandola, L.~J., Bauschlicher, C.~W., Boogert, A.~C.~A., 
Hayward, T.~L., Hudgins, D.~M., \& Sandford, S.~A.\ 2003, Astrophysics of 
Dust p.42

\bibitem[Peeters et al.(2005)]{2005SSRv..119..273P} Peeters, E., 
Mart{\'{\i}}n-Hern{\'a}ndez, N.~L., Rodr{\'{\i}}guez-Fern{\'a}ndez, N.~J., 
\& Tielens, X.\ 2005, Space Science Reviews, 119, 273 
\bibitem[Povich et al.(2007)]{2007ApJ...660..346P} Povich, M.~S., et al.\ 
2007, \apj, 660, 346 

\bibitem[aa]{aa} Price, S.D., Egan, M.P., Carey, S.J., Mizuno, D., Kuchar, T.
  2001, A\&A, 121, 2819

\bibitem[Rieke et al.(2004)]{2004ApJS..154...25R} Rieke, G.~H., et al.\ 
2004, \apjs, 154, 25 

\bibitem[Robitaille et al.(2006)]{2006ApJS..167..256R} Robitaille, T.~P., 
Whitney, B.~A., Indebetouw, R., Wood, K., \& Denzmore, P.\ 2006, \apjs, 
167, 256 

\bibitem[Robitaille et al.(2007)]{2007ApJS..169..328R} Robitaille, T.~P., 
Whitney, B.~A., Indebetouw, R., \& Wood, K.\ 2007, \apjs, 169, 328 
\bibitem[Smith et al.(2006)]{HS06} Smith, H. A., Hora, J. L., Marengo, M., \& Pipher, J. L. 2006 \apj, 645, 1264 
\bibitem[Shepherd et al.(2007)]{SP07} Shepherd, D. S. et al.\ 2007,
  \apj, submitted

\bibitem[Sohn et al.(2006)]{2006A&A...445...69S} Sohn, Y.-J., Kang, A., 
Rhee, J., Shin, M., Chun, M.-S., \& Kim, H.-I.\ 2006, \aap, 445, 69 

\bibitem[aa]{aa}Weaver, R., McCray, R., Castor, J., Shapiro, P., Moore, R. 1977, ApJ, 218, 377

\bibitem[Werner et al.(2004)]{2004ApJS..154....1W} Werner, M.~W., et al.\ 
2004, \apjs, 154, 1 

\bibitem[Whitney et al.(2003a)]{WI} Whitney, B. A., Wood, K.,
  Bjorkman, J. E., \& Wolff, M. J. 2003, \apj, 591, 1049

\bibitem[Whitney et al.(2003b)]{2003ApJ...598.1079W} Whitney, B.~A., Wood, 
K., Bjorkman, J.~E., \& Cohen, M.\ 2003, \apj, 598, 1079 

\bibitem[Whitney et al.(2004)]{2004ApJ...617.1177W} Whitney, B.~A., 
Indebetouw, R., Bjorkman, J.~E., \& Wood, K.\ 2004, \apj, 617, 1177 

\bibitem[aa]{aa}Whitworth, A.P., Bhattal, A.S., Chapman, S.J., Disney, J.J. \& Turner,
J.A. 1994, MNRAS, 268, 291

\bibitem[aa]{aa}Vink, J.S., deKoter, A., Lamers, H.J.G.L. 2001, A\&A, 369, 574
\end{thebibliography}
\end{document}